\documentclass[]{aa}
\usepackage{graphicx, natbib, epsfig, amsmath, amssymb, mathrsfs, txfonts}
\DeclareGraphicsExtensions{.eps, .jpg, .ps}
\usepackage{epstopdf}
\bibpunct{(}{)}{;}{a}{}{,}
\begin{document}

\title{The self-coherent camera as a focal plane fine phasing sensor}

\author{P. Janin-Potiron \inst{1}\and P. Martinez \inst{1}\and P. Baudoz \inst{2}\and M. Carbillet \inst{1}}

\institute{Laboratoire Lagrange, UMR7293, Universit\'e de Nice Sophia-Antipolis, CNRS, Observatoire de la C\^ote d\'{}Azur, Parc Valrose, B\^at. Fizeau, 06108 Nice Cedex 2, France
\and LESIA, Observatoire de Paris, CNRS and University Denis Diderot Paris 7, 5 place Jules Janssen, 92195 Meudon, France}

\offprints{Pierre.Janin-Potiron@oca.eu}

\abstract
{Direct imaging of Earth-like exoplanets requires very high contrast imaging capability and high angular resolution. Primary mirror segmentation is a key technological solution for large-aperture telescopes because it opens the path toward significantly increasing the angular resolution. 
The segments are kept aligned by an active optics system that must reduce segment misalignments below tens of nm RMS to achieve the high optical quality required for astronomical science programs.} 
{The development of cophasing techniques is mandatory for the next generation of space- and ground-based segmented telescopes, which both share the need for increasing spatial resolution. We propose a new focal plane cophasing sensor that exploits the scientific image of a coronagraphic instrument to retrieve simultaneously piston and tip-tilt misalignments.}
{The self-coherent camera phasing sensor (SCC-PS) adequately combines the SCC properties to segmented telescope architectures with adapted segment misalignment estimators and image processing. An overview of the system architecture, and a thorough performance and sensitivity analysis, including a closed-loop efficiency, are presented by means of numerical simulations.}
{The SCC-PS estimates simultaneously piston and tip-tilt misalignments and corrects them in closed-loop operation in a few iterations.
As opposed to numerous phasing sensor concepts the SCC-PS does not require any a priori on the signal at the segment boundaries or any dedicated optical path. We show that the SCC-PS has a moderate sensitivity to misalignments, virtually none to pupil shear, and is by principle insensitive to segment gaps and edge effects. Primary mirror phasing can be achieved with a relatively bright natural guide star with the SCC-PS.}
{The SCC-PS is a noninvasive concept and an efficient phasing sensor from the image domain. It is an attractive candidate for segment cophasing at the instrument level or alternatively at the telescope level, as usually envisioned in current space- and ground-based observatories.}


\authorrunning{Janin-Potiron et al.}

\maketitle

\section{Introduction}

Segmented telescopes are considered as the next leap forward for ground- and space-based observations. Their optical characteristics will deliver unreached high resolution and sensitivity for new thrilling science observations such as exoplanets direct imaging.
Segmented mirrors are an essential and unavoidable feature for the extremely large telescopes (ELTs), and will be necessary for the future exoplanet space-based projects for increasing telescope size despite the limited room offered within a given launch vehicle. For instance, the James Webb Space Telescope (JWST) will offer unprecedented spatial resolution by featuring a 6.5-meter segmented primary mirror. Its potential successor, the large UV/optical IR surveyor (LUVOIR), identified in the NASA long-term astrophysics road map, is expected to provide a larger segmented aperture (8 to 16\,m). 

Nonetheless, primary mirror segmentation highly complicates the structure of the pupil.
Tackling the various effects related to segmentation to achieve a certain level of image quality is fundamental. For the piston case only, the Strehl ratio depends exponentially on the RMS aberrations of the pupil \citep{YAITSKOVA03, TROY06}.
In particular, the most well-known problem is that of achieving a smooth continuous mirror surface, which is a process known as cophasing. The alignment of segments is mandatory as their misalignments correspond to stochastic effects (the value of characteristics changes randomly from segment to segment) and produce speckles in the scientific image. A properly phased telescope has a resolution that is comparable to the total diameter of the entire segmented primary mirror, whereas an unphased telescope delivers a very poor spatial resolution that is limited by the diameter of an individual segment.
Telescope cophasing usually requires two dedicated sensors (coarse and fine) to correct from the initial misalignment imposed by mechanical structure constraints to the final alignment (e.g., from $\sim$100\,$\mu$m to $\sim$10\,nm in the visible spectral range).

In this context, several methods have been proposed to measure fine segment misalignments, where various cophasing sensors are based on existing wavefront sensors usually employed in adaptive optics (AO) applications.
Cophasing methods can be divided into three flavors: (1) pupil plane techniques, such as the modified Shack-Hartmann \citep[e.g., ][]{CHANAN89, CHANAN00, MAZZOLENI08}, the Mach-Zehnder interferometer \citep{MONTOYATHESIS, YAITSKOVA04, YAITSKOVA05}, and avatars \citep{DOHLEN06, SURDEJTHESIS}; (2) intermediate plane techniques such as the curvature sensor \citep[e.g.,][]{CHANAN99, CUEVAS00, CHUECA08}; and (3) focal plane techniques such as the phase diversity \citep{LOFDAHL98,DELAV10}, Pyramid sensor \citep{ESPOSITO05, PINNA08}. 
More recent and peculiar approaches have been investigated using pupil asymmetry \citep[asymmetric pupil wavefront sensor,][]{MARTINACHE13, POPE2014} or differential optical transfer functions \citep{CODONA15}.

Notwithstanding these various developments, high performance and image quality require the primary mirror segments to be aligned at the level of a few tens of nanometers for most applications, but the requirements of high-precision phasing for high contrast imaging instruments is still uncertain and might be much more stringent. In particular, ELTs and potentially the next generation of space-based observatories represent a significant change in dimension, wavefront control strategies, and execution time.
Including target acquisition, calibrations, closed-loop iterations, and computation time, phasing an ELT from coarse-to-fine and fine-to-final phasing levels would require about one to two hours operation with state-of-the-art phasing sensors.
In the present paper, we propose a new concept based on the self-coherent camera principle \citep[SCC;][]{BAUDOZ06} for the purpose of fine cophasing.
The self-coherent camera phasing sensor (SCC-PS) adequately combines the SCC properties to segmented telescopes with adapted segment misalignment estimators and image processing and has several advantages in comparison with other techniques.
It directly exploits the scientific image to disentangle and retrieve unambiguously and simultaneously segment piston and tip-tilt misalignments.
This low cost sensor is elegant for its simplicity and benefit in terms of implementation, alignment, calibration, accuracy, and computation time.
Ultimately, it opens the path toward real-time fine-tuning phasing and temporal stability control during on-sky observation. 
This paper presents an overview of the SCC-PS and provides a thorough performance and sensitivity analysis of the system, including a closed-loop efficiency that is reported by means of numerical simulations. 

The paper reads as follows: in Sect. \ref{subsec:principle} we present the principle of the SCC-PS, the estimators to measure piston, tip, and tilt of each segment and discuss the capture range of the sensor.
In Sect. \ref{Numerical} we provide details about several aspects, such as the system calibration, system architecture, and parameter space. In Sect. \ref{sec:Results} we discuss the performance of the SCC-PS including a sensitivity analysis to misalignments, close-loop efficiency, and sky coverage. 
Finally, we draw conclusions and list a set of advantages of the SCC-PS over the state-of-the-art phasing sensors in Sect. \ref{conclusion}.

\section{Principle and estimators}
\label{subsec:principle}

\begin{figure*}[!ht]
\centering
\includegraphics[width=\linewidth]{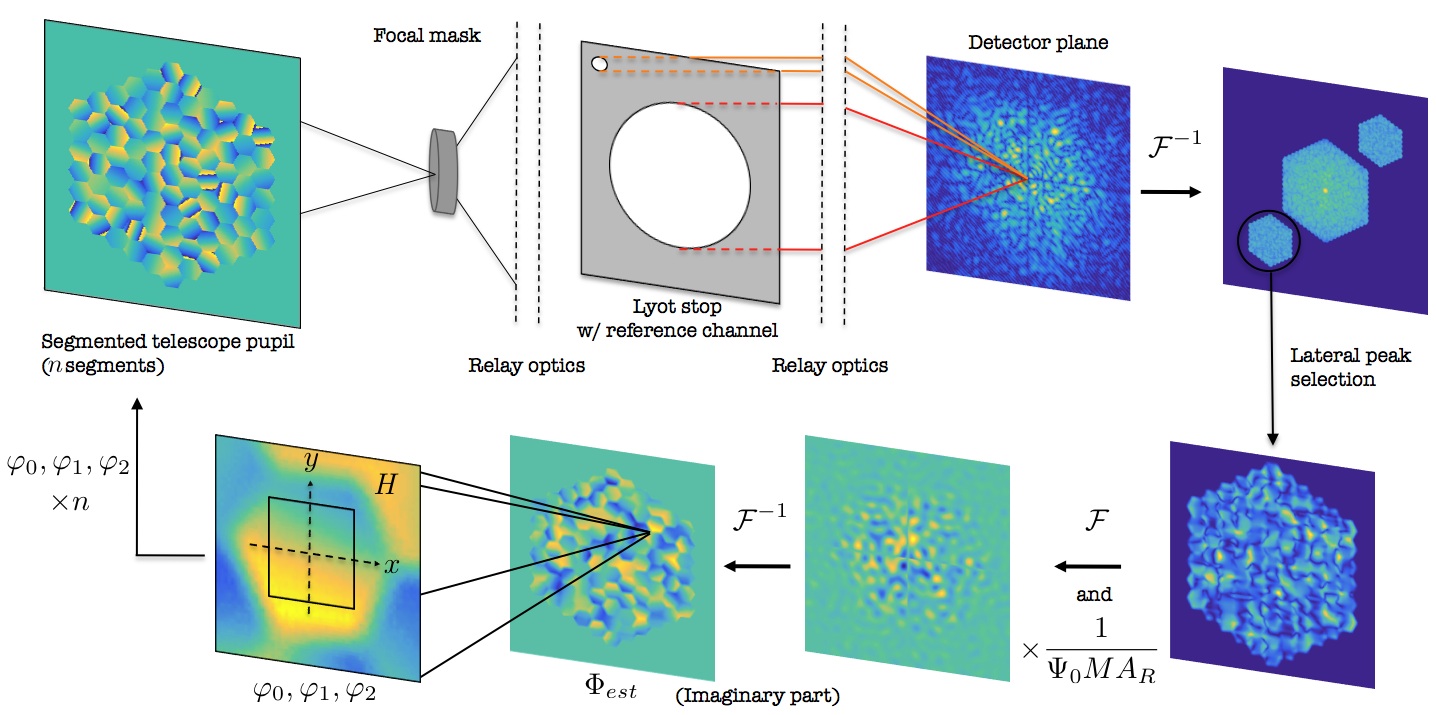}
\caption{Illustration of the self-coherent camera phasing sensor (SCC-PS) principle. The red rays correspond to the path of the light in a classical coronagraph propagation process. The orange rays correspond to the path of light in the reference pupil in the Lyot stop. From the estimated phase map ($\phi_{est}$) three estimators are evaluated on each segment ($\varphi_0$, $\varphi_1$, and $\varphi_2$). }
\label{SchemaPrincipe}
\end{figure*}

The SCC \citep{BAUDOZ06,GALICHER08} uses the coherence of light to generate Fizeau fringes in the image plane to spatially encode speckles. The SCC exploits the coherence of light 
between the post-coronagraphic stellar speckles and the stellar light rejected by a coronagraph outside from the relayed geometrical pupil.
By combining appropriate estimators and the SCC, one can correct for instrumental aberrations (quasi-static errors) upstream of the coronagraph, and the SCC transforms into an efficient focal plane wavefront sensor \citep{MAZOYER13}. Such a system can be used in a closed-loop manner with a deformable mirror (DM) in the pupil plane of a scientific instrument to correct for both phase and amplitude aberrations. In this section, we briefly review the formalism of the SCC, and describe how this instrument can be used as a focal plane wavefront sensor for cophasing purpose using adequate segment misalignment estimators and image processing. For the sake of clarity, we follow the formalism used in \citet{MAZOYER13}. 
In the theoretical description hereafter, we assume a monochromatic light coming from an unresolved object using Fourier optics. The general scheme of the SCC-PS principle is presented in Fig. \ref{SchemaPrincipe}, where 
the primary mirror of the telescope is segmented and arbitrarily composed of 91 segments. For the sake of similarity with \citet{MAZOYER13}, the coronagraph used is the four quadrant phase mask \citep[FQPM;][]{ROUAN00}, whereas the SCC-PS can accommodate various coronagraph types or even noncoronagraphic but purely diffractive components. The stellar light is assumed to be ideally centered on the optical axis and, in particular, no pointing errors or pupil shear are considered.

\subsection{General analytical description}
\label{subsec:analytical}

Assuming only phase aberrations on the incoming wavefront, the complex amplitude in the segmented entrance pupil plane can be expressed as
\begin{equation}
\psi'_S(\boldsymbol{\xi},\lambda) = P(\boldsymbol{\xi}) \times \psi_0 \exp(i\phi( \boldsymbol{\xi} ,\lambda) ),
\label{PsiPrime}
\end{equation}
where $\boldsymbol{\xi}$ is the spatial coordinate in the pupil plane, $\psi_0$ the constant amplitude of the electrical field, and $\phi(\boldsymbol{\xi})$ represents the wavefront phase aberration. The parameter $P(\boldsymbol{\xi})$ is used here to define the pupil limits (unity value inside the pupil and zero elsewhere).

Assuming that the aberrations are small enough and defined over the pupil (i.e., $P\phi = \phi$), Eq. \ref{PsiPrime} can be simplified as
\begin{equation}
\frac{\psi'_S}{\psi_0} = P(\boldsymbol{\xi}) + i\phi( \boldsymbol{\xi} ).
\label{PsiPrime2}
\end{equation}
The light then focalizes onto a coronagraphic or diffractive mask. 
The complex amplitude $A'_S$ behind the coronagraphic mask \textit{M} is given by
\begin{equation}
A'_S = \mathcal{F}[\psi'_S]M,
\label{APrime}
\end{equation}
where $\mathcal{F}$ stands for the Fourier transform. 
The expression of the complex amplitude after the classical Lyot stop (\textit{L}) is given by
\begin{equation}
\frac{\psi_S}{\psi_0} = \frac{\mathcal{F}^{-1}[A'_S]}{\psi_0} L= \left [ (P + i \phi) \ast \mathcal{F}^{-1}[M] \right ] L,
\label{Psi}
\end{equation}
where $\ast$ represents the convolution product.
Assuming that the nonaberrated part of the electric field is null inside the imaged pupil, the complex amplitude in the focal plane can be expressed by
\begin{equation}
A_S = \mathcal{F}[\psi_S] = i\psi_0 (\mathcal{F}[\phi] M) \ast \mathcal{F}[L].
\label{As}
\end{equation}
As shown in Eq. \ref{As}, \textit{$A_S$} is directly dependent on the aberrations $\phi$ in the entrance pupil. At this point we have no direct access to the complex amplitude but the intensity given by the squared modulus of the electrical field,
\begin{equation}
I = \lvert A_S \rvert ^2 \propto \mathcal{F}[\phi] \mathcal{F}[\phi]^{\ast},
\label{PsiM}
\end{equation}
and as a consequence the information needed to invert Eq. (\ref{As}) to retrieve the function $\phi$ is lost.
To tackle this limitation \citet{BAUDOZ06} proposed the concept of the SCC that was originally build as an interferometer, whereas \citet{GALICHER10} proposed an elegant and robust design by associating the SCC with a coronagraph using a modified Lyot stop. It basically consists of adding a small noncentered hole (the reference channel, $R$ hereafter) into the Lyot stop mask (see Fig. \ref{SchemaPrincipe}). 
This small hole is inserted outside the imaged pupil where the coronagraph has rejected the stellar light, and the light passing through the Lyot stop but the reference hole is the pupil channel. 
\noindent From Eq. (\ref{APrime}), the electric field after the modified Lyot stop can be formalized as
\begin{equation}
\begin{split} 
\frac{\psi(\boldsymbol{\xi},\lambda)}{\psi_0} = \left [ ( P(\boldsymbol{\xi}) + i\phi(\boldsymbol{\xi},\lambda) ) \ast \mathcal{F}^{-1}[M](\boldsymbol{\xi}) \right ] \\ ~ \times ~ (L(\boldsymbol{\xi}) + R(\boldsymbol{\xi})\ast \delta(\boldsymbol{\xi} - \boldsymbol{\xi}_0)),
\label{PsiModifiedLyot}
\end{split} 
\end{equation}
where $\boldsymbol{\xi}_0$ is the separation between the pupil and reference channel in the Lyot stop and $\delta$ stands for the Dirac delta function. The equation simplifies as
\begin{equation}
\psi(\boldsymbol{\xi},\lambda) = \psi_S(\boldsymbol{\xi},\lambda) + \psi_R(\boldsymbol{\xi},\lambda) \ast \delta(\boldsymbol{\xi} - \boldsymbol{\xi}_0),
\label{PsiModifiedLyot2}
\end{equation}
where $\psi_S$ is the complex amplitude in the pupil channel (classical Lyot stop) as defined in Eq. (\ref{Psi}) and $\psi_R$ in the reference channel. The intensity in the image plane can be expressed by calculating $I = \lvert \mathcal{F}(\psi) \rvert ^2$ that is given by
\begin{equation}
\begin{split}
I(\boldsymbol{x}) = \lvert A_S(\boldsymbol{x}) \rvert ^2 + \lvert A_R(\boldsymbol{x}) \rvert ^2 + A_S^{\ast}(\boldsymbol{x}) A_R(\boldsymbol{x}) \exp \left (-\frac{2i\pi\boldsymbol{x}.\boldsymbol{\xi}_0}{\lambda} \right ) \\ + A_S(\boldsymbol{x}) A_R^{\ast}(\boldsymbol{x}) \exp \left (\frac{2i\pi\boldsymbol{x}.\boldsymbol{\xi}_0}{\lambda} \right ), 
\end{split}
\label{EqSCC}
\end{equation}
where $A_S$ and $A_R$ represent the Fourier transform of $\psi_S$ and $\psi_R$, respectively, and $\boldsymbol{x}$ represents the coordinate in the image plane. Fringes are formed by the interference between the on-axis and diffracted light from the star, whereas the planet is not affected by the coronagraph. The modulations of the fringes are directly related to the complex amplitudes $A_S$ and $A_R$.

\subsection{Phase retrieval, piston, and tip-tilt estimators}
\label{subsec:PhaseRetrieval}

We present the principle of the complex amplitude estimation from the speckle field obtained with the SCC-PS to retrieve segment misalignment information. 
Applying an inverse Fourier transform to Eq. (\ref{EqSCC}) leads to
\begin{equation}
\begin{split}
\mathcal{F}^{-1}[I](u) = \mathcal{F}^{-1}[I_S + I_R] + \mathcal{F}^{-1}[ A_S^{\ast}(\boldsymbol{x}) A_R(\boldsymbol{x}) ] \ast \delta(\boldsymbol{u} - \frac{\boldsymbol{\xi_0}}{\lambda}) \\ + \mathcal{F}^{-1}[ A_S(\boldsymbol{x}) A_R^{\ast}(\boldsymbol{x}) ] \ast \delta(\boldsymbol{u} + \frac{\boldsymbol{\xi_0}}{\lambda}),
\label{InverseSCC}
\end{split}
\end{equation}
where $I_S= \lvert A_S \rvert ^2$ and $I_R= \lvert A_R \rvert ^2$ are the intensities of the speckles obtained from the pupil and reference channels, respectively, as demonstrated in \citet{MAZOYER13}, and $\boldsymbol{u}$ is the coordinate in the Fourier plane.
The image obtained after this inversion is composed of three entities centered in $\boldsymbol{u} = [-\xi_0 / \lambda,0,\xi_0 / \lambda]$ as shown in Fig. \ref{SchemaPrincipe} (top right corner image). The two lateral peaks are referred to as $\mathcal{F}^{-1}[I_-]$ and $\mathcal{F}^{-1}[I_+]$.

We denote $D_L$ the diameter of the pupil aperture (circumscribed diameter, i.e., diameter of pupil channel) and $D_R$ the diameter of the reference channel. For convenience, we define the ratio parameter $\gamma$ of these two quantities as follows:
\begin{equation}
\gamma = \frac{D_L}{D_R} \implies D_R = \frac{D_L}{\gamma}.
\label{def_gamma}
\end{equation}

As demonstrated in \cite{GALICHER10} for circular apertures, the reference channel has to satisfy a condition on its position, as the central and lateral peaks would be misidentified otherwise.
In the case of an hexagonal pupil, the symmetry of the problem is slightly different.

The condition can be expressed as
\begin{equation}
\lvert \lvert \boldsymbol{\xi_0} \rvert \rvert > \frac{D_L}{2\cos(\theta)} \left ( 3+\frac{1}{\gamma} \right ) \text{~~~with~~~} -\frac{\pi}{6} < \theta < \frac{\pi}{6} ~~ \left [ \frac{\pi}{3} \right ],
\label{condition}
\end{equation}
where $\theta$ is the angle between the horizontal and the reference.
Taking the last term of Eq. \ref{InverseSCC} and centering it in $\boldsymbol{u} = 0$, we obtain
\begin{equation}
\mathcal{F}^{-1} [ I_-] = \mathcal{F}^{-1} [ A_S A_R^{\ast}].
\label{Im}
\end{equation}
Assuming $\gamma \gg 1$, we can consider the complex amplitude in the reference channel to be small enough to consider $A_R^{\ast}$ uniform over the correction zone. Under these assumptions, Eq. \ref{Im} gives
\begin{equation}
A_S = \frac{I_-}{A_R^{\ast}}.
\label{As2}
\end{equation}
Combining Eq. \ref{As} and Eq.\ref{As2}, \citet{MAZOYER13} proposes an estimator for both phase and amplitude aberrations. Considering phase aberrations only it yields
\begin{equation}
\phi_{est} = \Im \left( g \mathcal{F}^{-1} \left [ \frac{I_-}{M} \right ] P \right),
\label{PhiEst}
\end{equation}
where $\Im(z)$ is the imaginary part of $z$ and $g$ is a constant that can be adjusted like a gain.
The size of the reference channel plays a significant role in the accuracy of the phase measurement. 
As seen in Eq. \ref{Im},
\begin{equation}
\phi_{est} \propto \mathcal{F}^{-1}[A_S] \ast \mathcal{F}^{-1}[A_R^{\ast}],
\label{PhiEstprop}
\end{equation}
and the larger the reference channel the lower the resolution for accurate estimation of the phase due to the convolution by the Fourier transform of the reference. This point is developed further in Sect. \ref{criterion}.
Figure \ref{SchemaPrincipe} summarizes the steps required for the estimation of the segment misalignments with the SCC-PS.

To retrieve and disentangle the piston and tip-tilt information from $\phi_{est}$, we need to define adequate estimators. 
In contrast to various phasing methods \citep[e.g., the Mach-Zehnder phasing sensor and avatars;][]{YAITSKOVA04, SURDEJ10}, measurement across the intersegment is irrelevant as the information can be extracted directly from the segment surface through the wavefront error map obtained with the SCC-PS. Therefore, the SCC-PS does not require any a priori knowledge of the theoretical signal at the segment boundaries. The SCC-PS is therefore insensitive to segment gaps or edge effects. 
Piston and tip-tilt information of individual segments correspond to a vertical translation on $\boldsymbol{z}$-axis, rotation around $\boldsymbol{x}$-axis, and rotation around $\boldsymbol{y}$-axis, respectively, and can be obtained using simple metrics applied to $\phi_{est}$. 
As the SCC-PS delivers a copy of the phase distribution in the entrance plane literally, we can define trivially three estimators, $\varphi_{0}$, $\varphi_{1}$, and $\varphi_{2}$, for piston, tip, and tilt, respectively. 
For each segment we define a square zone (\textit{H}) of side length $h$ centered on the segment position where the signal for the integrals defined hereafter are estimated.
 
For the estimation of the piston we calculate the integral of the signal $\phi_{est}$ over the surface defined by \textit{H} and $\varphi_0$ is expressed as
\begin{equation}
\varphi_0 = \iint\limits_H \phi_{est} \mathrm{d}x \mathrm{d}y.
\label{EstPiston}
\end{equation}
For the estimation of the tip and tilt, the gradient of the signal $\phi_{est}$ is estimated following the $\boldsymbol{x}$- and $\boldsymbol{y}$-axis, respectively, and the estimators $\varphi_1$ and $\varphi_2$ are given by 
\begin{equation}
\varphi_1 = \iint\limits_H \nabla_x \left( \phi_{est} \right) \mathrm{d}x \mathrm{d}y,
\label{EstTip}
\end{equation} 
\begin{equation}
\varphi_2 = \iint\limits_H \nabla_y \left( \phi_{est} \right) \mathrm{d}x \mathrm{d}y,
\label{EstTip2}
\end{equation}
where $\nabla_x$ and $\nabla_y$ stand for the gradient in the direction $x$ and $y,$ respectively.

\subsection{Capture range}

\begin{figure*}[!ht]
\centering
\includegraphics[height=0.38\textwidth]{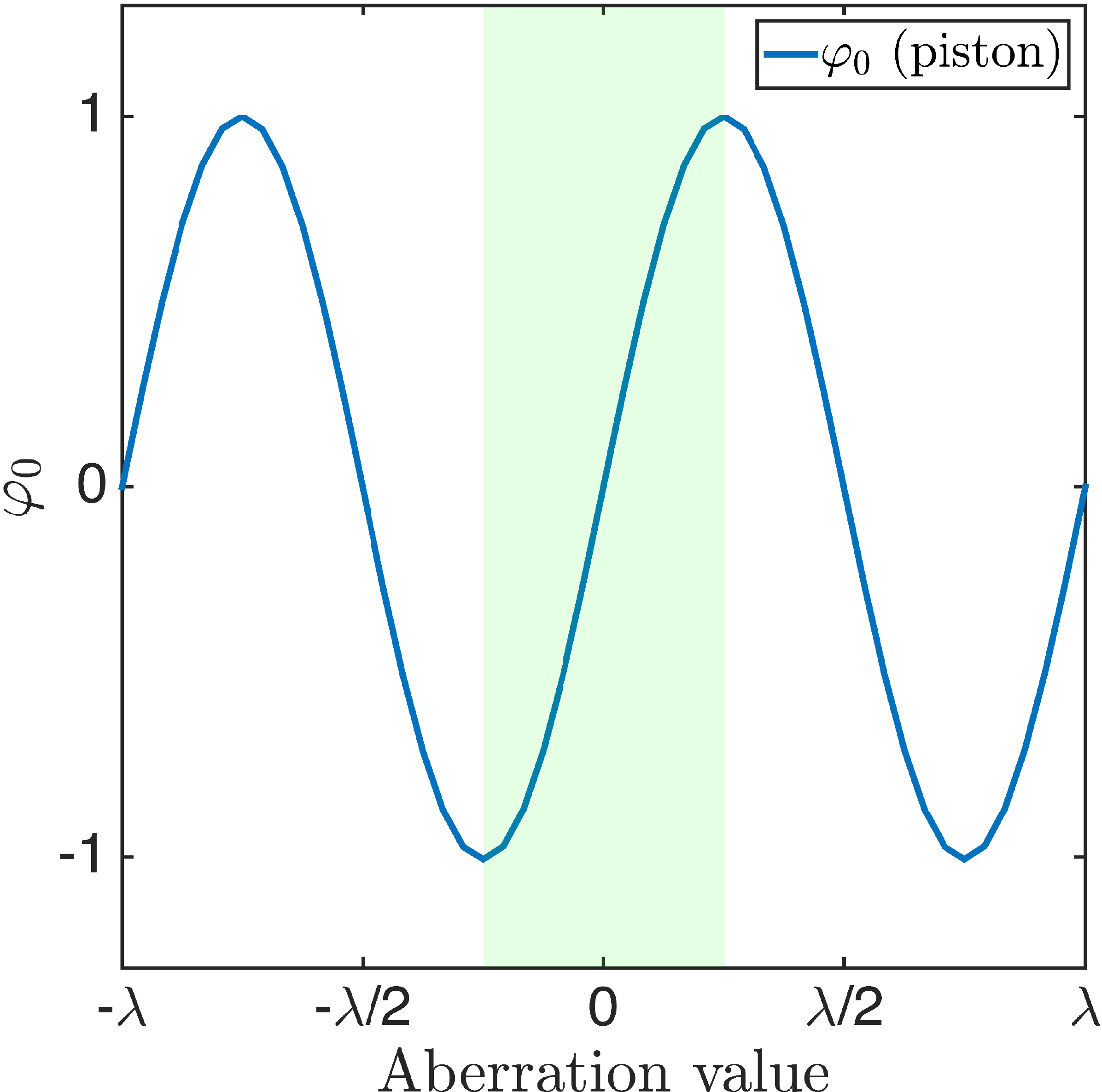}
\hspace{1cm}
\includegraphics[height=0.38\textwidth]{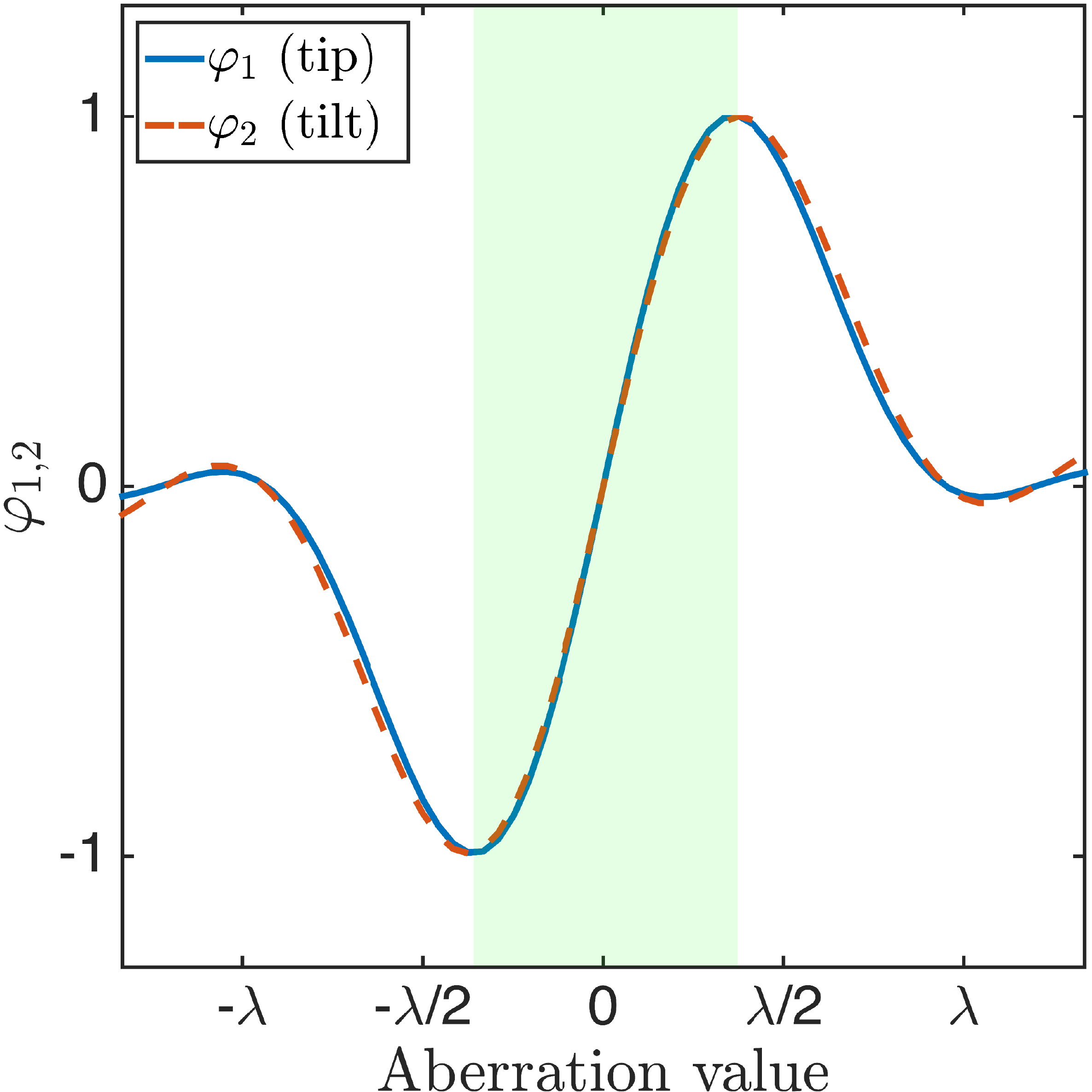}
\caption{Representation of measured estimators (normalized) as a function of the introduced aberration value for piston (left) and for tip-tilt (right). Green zones represent the capture range of the sensor. In this case, $h \simeq 23\%$ of the flat-to-flat distance of the segment}
\label{PistonEstimator}
\end{figure*}

\begin{figure*}[!ht]
\centering
\includegraphics[height=0.4\textwidth]{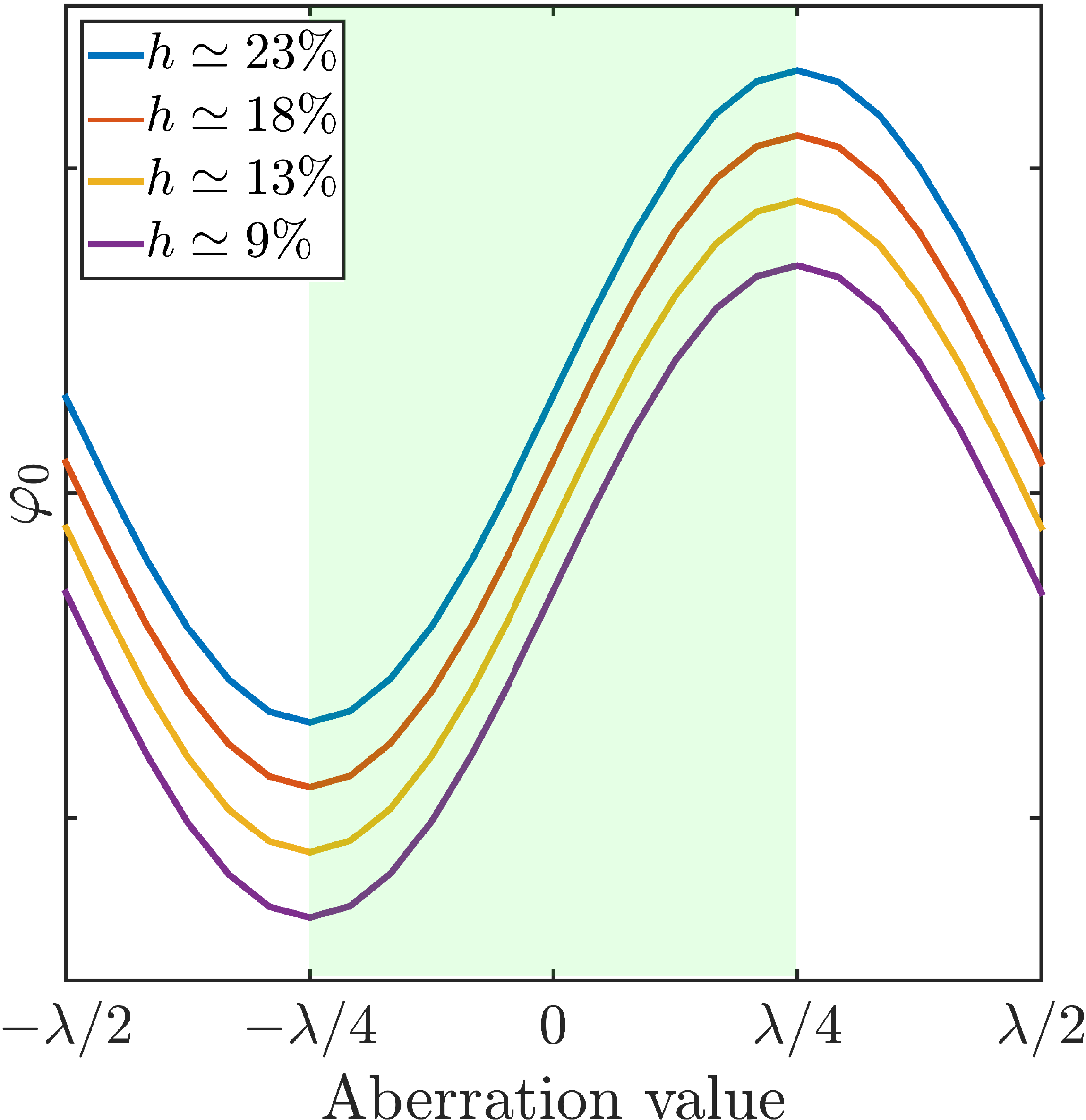}
\hspace{1cm}
\includegraphics[height=0.4\textwidth]{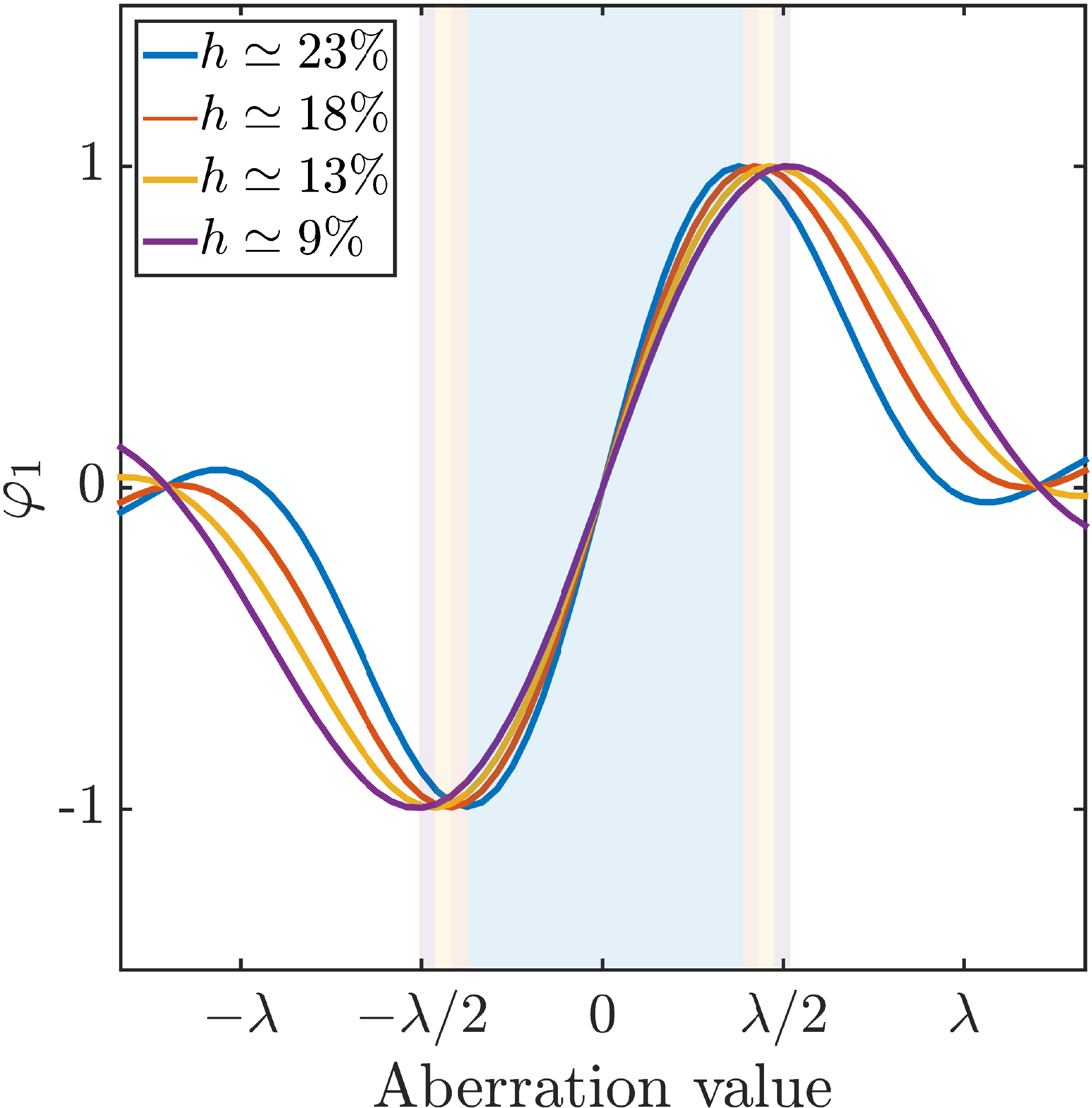}
\caption{Influence of the $H$-zone width ($h$ given in percentage of the segment flat-to-flat distance) on the reachable capture range when considering piston only (left) and tip only (right). For piston only, otherwise superimposed, all curves have been arbitrary shifted with different offsets in the y-axis to disentangle them.}
\label{TtEstimatorZone}
\end{figure*}

The theoretical treatment presented previously describes the properties of the SCC-PS for cophasing performance within the single-wavelength capture range.
The capture range of each individual estimator of the SCC-PS is assessed from $-\lambda$ to $\lambda$ and is presented in Fig. \ref{PistonEstimator}.
For the piston (Fig. \ref{PistonEstimator}, left), the estimated phase is as expected $\lambda$-periodic and the limitation in the reachable bijective zone is as discussed for instance in \cite{VIGAN11} limited from $-\lambda/4$ to $\lambda/4$, the so-called $\pi$-ambiguity problem that appears in all phasing sensor operating in monochromatic light. Outside this capture range, the position measurements are incorrect and lead to a situation in which some segments are shifted by the integer of the wavelength instead of being phased to the zero step. 
The tip-tilt estimator exhibits a larger capture range from $-3\lambda/8$ to $3\lambda/8$ (Fig. \ref{PistonEstimator}, right).
The capture range of the SCC-PS in monochromatic light is thus considered limited from $-\lambda/4$ to $\lambda/4$.
Within the capture range, the phase (piston and tip-tilt) can be reconstructed unambiguously.
 Changing the size where we apply the integral for evaluating the tip and tilt estimates, $\varphi_1$ and $\varphi_2$, implies a moderate change in the length of the capture range. Figure 3 (right) shows that the smaller the signal zone the larger the capture range, where the effect might be due to the sinusoidal shape of the signal itself and the way the tip-tilt estimators are built. 
This effect is not perceivable with the piston, as shown in Fig. \ref{TtEstimatorZone} (left), because the signal is smooth and regular all over the segment.

The $\pi$-ambiguity is common to any phasing sensors. To overcome this problem and to increase the capture range of the phasing sensor, existing techniques can straightforwardly be applied to the SCC-PS such as the dual-wavelength and coherence methods \citep[e.g.,][]{VIGAN11, CHANAN98}.

The dual-wavelength method performs successive phasing at two different wavelengths using two optical filters ($\lambda_1$ and $\lambda_2$) to identify and correct the ambiguity for every segment. The measurable aberration span undertakes a step forward from $\pm\lambda/4$ to $\pm(\lambda_1+\lambda_2)^2/2(\lambda_1-\lambda_2)$. Adjusting the two wavelengths allows us to establish a balance between precision and reachable capture range as long as the ambiguity remains identical at the two selected wavelengths.

The coherence technique uses the coherence length in the broadband light of an optical filter with a finite spectral bandwidth. The coherence function between two segments decreases as the piston aberration increases, and by scanning over a wide range of piston one can determine the variation in the coherence function, where the maximum of the coherence allows us to identify and correct the ambiguity for every segment. The coherence method enhances the capture range from $\pm\lambda/4$ to $\pm L_c$, where $L_c$ is the coherence length of the optical filter. The bandwidth of the optical filter is selected to balance the new capture range and the precision of the measurement at the optimal position.

Using these techniques, the capture range of a phasing sensor can be extended from $\pm \lambda/4$ to a few tens of $\mu$m depending on the optical filters and method selected \citep[e.g.,][]{CHANAN98, VIGAN11, SURDEJTHESIS}. We will devise practical solutions to improve the efficiency of these techniques, for instance, by taking advantage of the multireference architecture of the SCC \citep[e.g.,][]{DELORME16} and developing brand new solutions for increasing the capture range of the SCC-PS. Results of these developments will be presented in a forthcoming paper and, thus, simulations and developments presented in the present paper are restricted to the study of the SCC-PS in the nonambiguity domain.

\section{Calibration and properties}
\label{Numerical}

\subsection{Numerical assumptions}

Simulations assume a segmented hexagonal entrance pupil composed of 91 segments of 50 pixels width (corner to corner) over 5 hexagonal rings without any central obscuration nor secondary mirror supports as shown in Fig. \ref{SchemaPrincipe}. We make use of simple Fraunhofer propagators between pupil and image planes, which are implemented as fast Fourier transforms (FFTs) generated with a Matlab code.
The telescope pupil is roughly 500 pixels in diameter and the matrix are $4096\times4096$ pixels. The resulting sampling in the focal plane is about 8 pixels per $\lambda/D,$ where $D$ is the entrance pupil diameter.
We assume a turbulence-free system (no dynamical aberrations) that is consistent with a spatial application, or representative of a ground-based application in which the presence of a first stage consisting of an extreme AO system has already and ideally corrected the dynamical turbulence. We also assume a system free of aberrations with a spacial frequency that is inferior to the segment size. We use a monochromatic light of wavelength $\lambda=600$\,nm.

\begin{figure*}[!ht]
\centering
\includegraphics[width=0.24\textwidth,trim=450 80 400 50,clip=true]{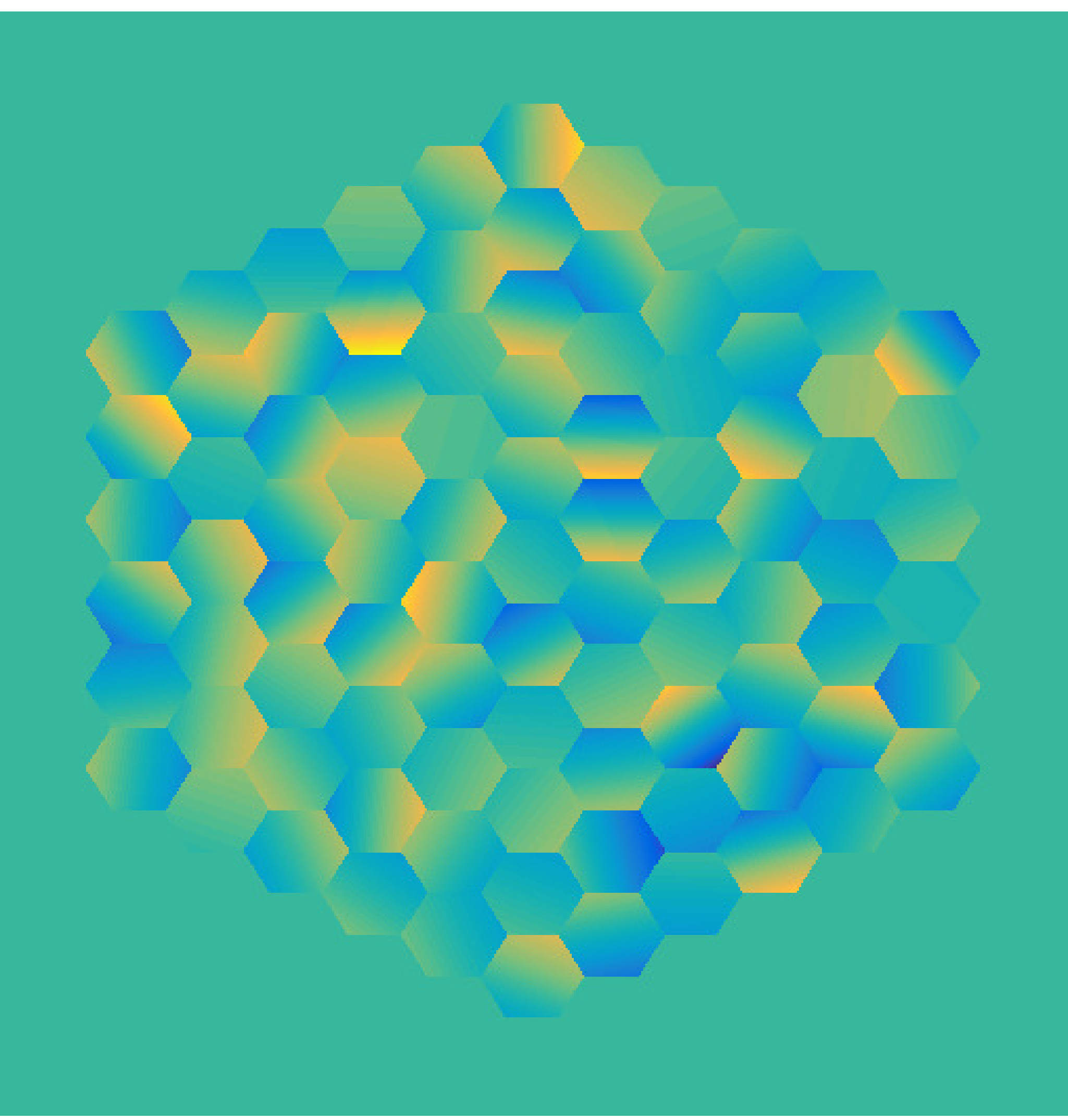}
\includegraphics[width=0.24\textwidth,trim=450 80 400 50,clip=true]{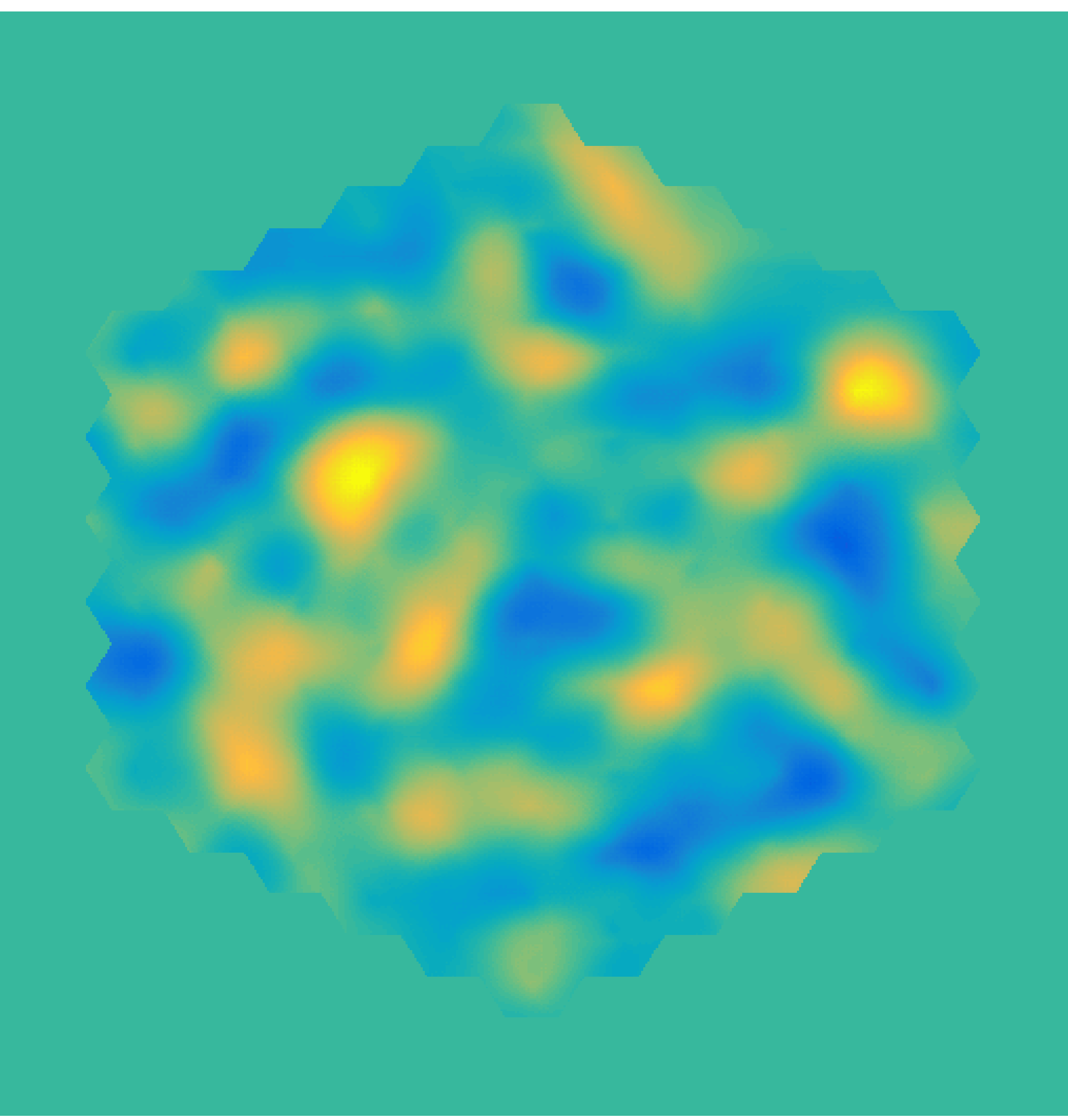}
\includegraphics[width=0.24\textwidth,trim=450 80 400 50,clip=true]{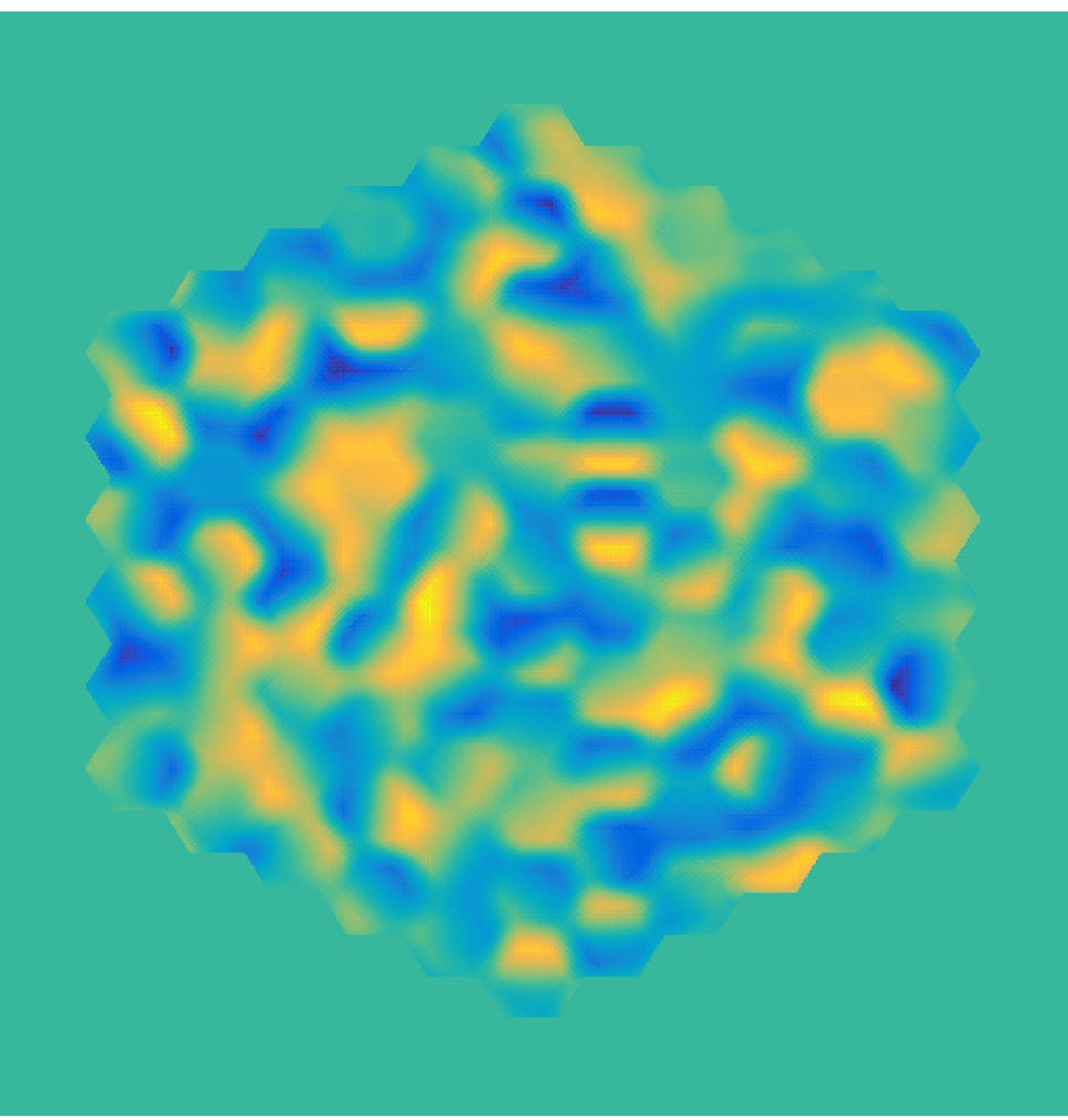}
\includegraphics[width=0.24\textwidth,trim=450 80 400 50,clip=true]{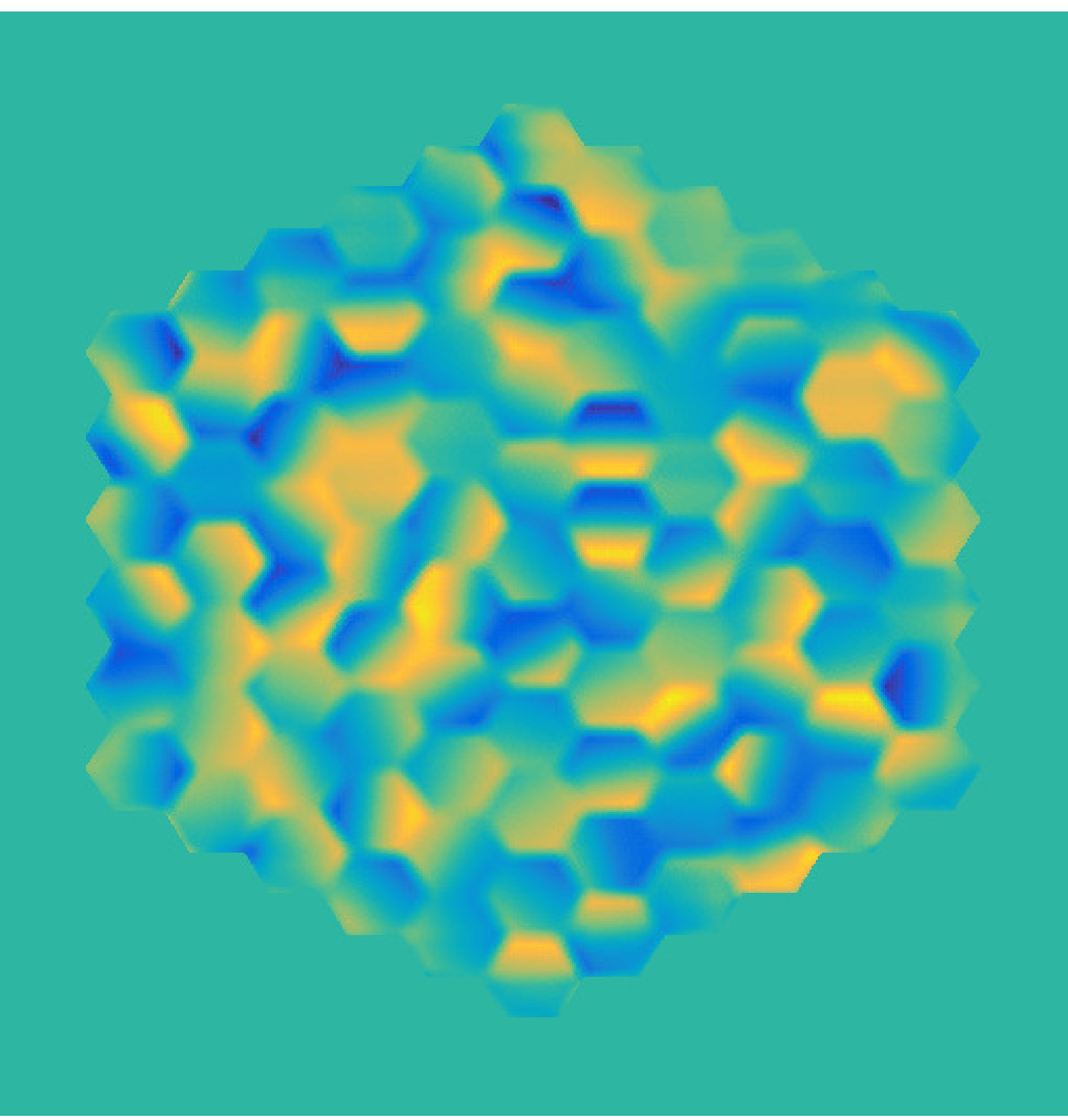}
\caption{Phase distribution (from left to right) in the entrance pupil and its estimated phase for different reference hole sizes: $\gamma = 10$; $\gamma = 30$; $\gamma = 50$.}
\label{ComparaisonGamma}
\end{figure*}
\subsection{Calibration of the system}
\label{subsec:calibration}

The SCC-PS calibration and operation aspects are similar to those of AO systems. The SCC-PS is calibrated using an interaction matrix, built by pushing each individual segment sequentially by a given value of piston, tip, and tilt, and by measuring the corresponding $\varphi_{0}$, $\varphi_{1} $, and $\varphi_{2}$ estimates (i.e., the system response). 
The amplitude of the pokes are chosen to ensure that the linear response of the system is respected. In practice we poke the segments by $\lambda/20$ in piston and by $\lambda/15$ in tip and tilt. 
The calibration matrix is obtained with a perfectly flattened mirror and during the calibrating process the system is perfectly aligned.
The calibration matrix contains dominant diagonal terms and relatively low values elsewhere. As we directly measure the estimators introducing a unique aberration at a time, and because the three estimators provide direct access to the three unknowns, the calibration matrix is square. 
Finally, the piston, tip, and tilt values of each individual segment are reconstructed by solving the system of linear equations using an inverted interaction matrix. At each iteration piston and tip-tilt estimates are applied as a correction to modify the shape of the primary mirror accordingly, and it iteratively converges toward an optimal mirror configuration.
As an optical quality metrics, we use the measurement of the residual RMS over the entire pupil and the Strehl ratio measured on the scientific image.

\subsection{Reference channel}
\label{Sec:Reference}

Three parameters play an important part in these simulations: (1) the position of the reference channel; (2) the flux in the reference channel; and (3) the size of the reference channel. 
As previously pointed out, the position of the reference channel must satisfy Eq. \ref{condition} to prevent information overlap in the Fourier domain.
Points (2) and (3) are developed hereafter.

\subsubsection{Flux equalization}
\label{Sec:Flux}

Equalization of fluxes between the pupil channel and reference channel is required as it directly impacts the contrast of the fringes.  
The amplitude gain required (mimicking the presence of a neutral density on the pupil channel) is calculated on a system with a perfectly flattened mirror and remains constant during the iterative correction process.
The value of this gain used in simulation is given by the square root of the energy ratio of the two channels (reference and pupil) and is defined as
\begin{equation}
\rho=\sqrt{(E_{R} / E_{S})},
\end{equation}
where $E_{R}$ is the energy inside the reference channel, and $E_{S}$ the energy in the pupil channel. 
The fringe contrast is optimum when the two fluxes are equal. We use in our simulation $3.10^{-3} \lesssim \rho \lesssim 6.10^{-3}$, depending on the configuration used.
\begin{figure}[!ht]
\centering
\includegraphics[height=0.38\textwidth]{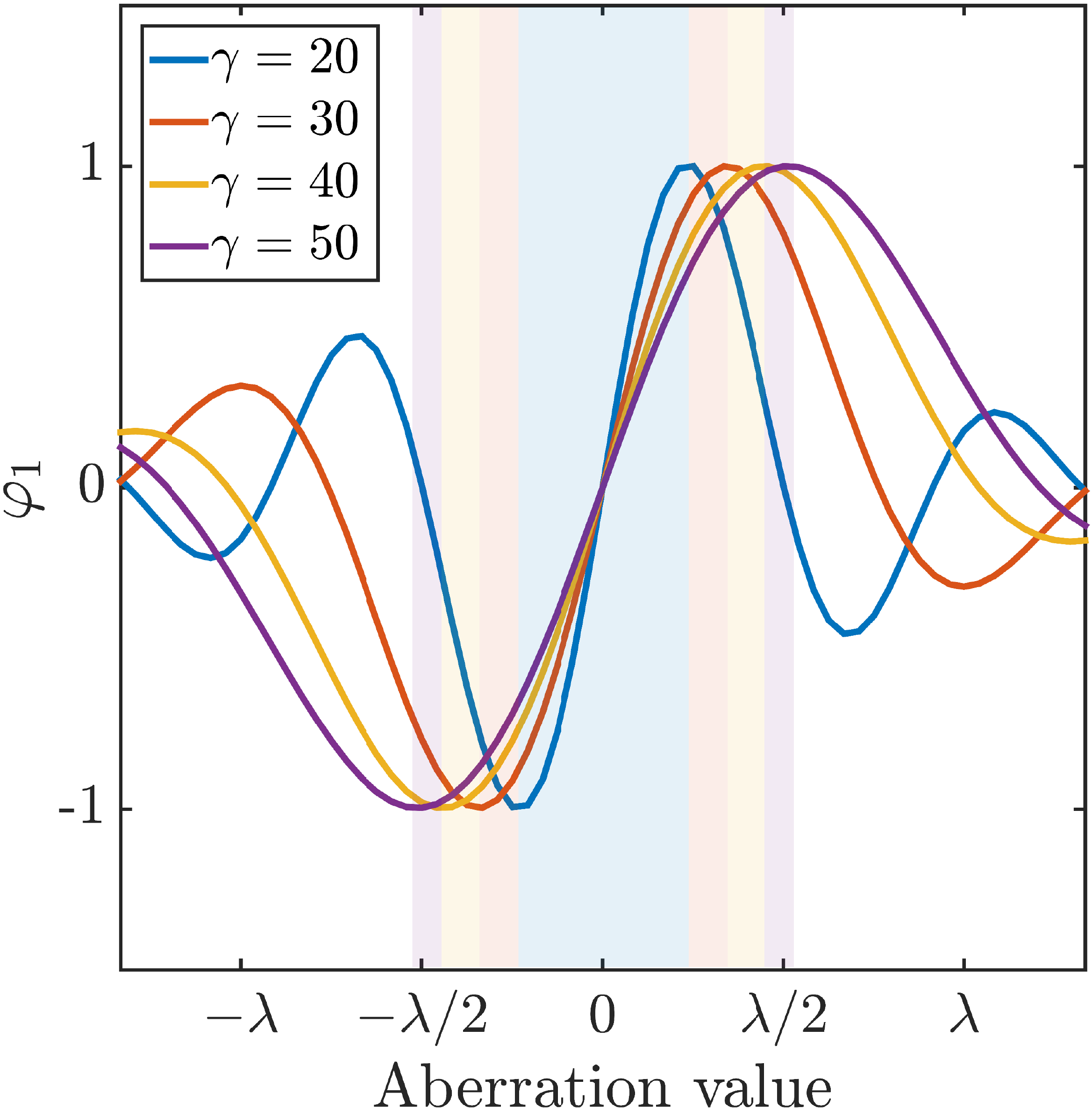}
\caption{Measured tip estimator (normalized) as a function of the introduced tip for different $\gamma$ values. Colored zones represent the respective capture range.}
\label{TipGamma}
\end{figure} 
\subsubsection{Size of the reference channel}
\label{criterion}
An adequate selection of the value of $\gamma$ (previously defined in Eq. \ref{def_gamma}) is needed as it influences the SCC-PS in two different ways: (1) it cannot be infinitely high because an appropriate signal-to-noise ratio (S/N) would not be guaranteed in the image plane otherwise; (2) but it must be high enough so that it allows retrieving the local phase associated with an individual segment.

Aspect (1) can be formalized following the initial formalism and by expressing the fringes contrast as
\begin{equation}
C= \frac{2 \sqrt{I_S I_R}}{ I_S + I_R },
\label{EQ:contraste}
\end{equation}
and when $I_R \ll I_S$ ($\iff \gamma \ggg 1$) the expression is simplified as
\begin{equation}
C \simeq 2 \sqrt{\frac{I_R}{I_S}} \ll 1.
\end{equation}
This situation is problematic because of inversion-related issues involved when noises occur on a real test bench as described in \cite{MAZOYER13}. 
In other words, the size of the reference channel influences the S/N on the fringes and thus modifies the reliability of the reference point-spread function as discussed in \citet{GALICHER10, MAZOYER13}. 

Regarding aspect (2), Fig. \ref{ComparaisonGamma} qualitatively illustrates that the higher the value of $\gamma$ the better the precision of the estimated phase. 
Because of the convolving operation of the phase signal with a hole of diameter inversely proportional to $\gamma$, the size of the reference channel directly drives the accuracy of $\varphi_{est}$ as seen in Eq. \ref{PhiEstprop}.
In addition, Fig. \ref{TipGamma} emphasizes the impact of $\gamma$ values on the capture range.

\begin{figure*}[!htp]
\centering
\includegraphics[height=0.38\textwidth]{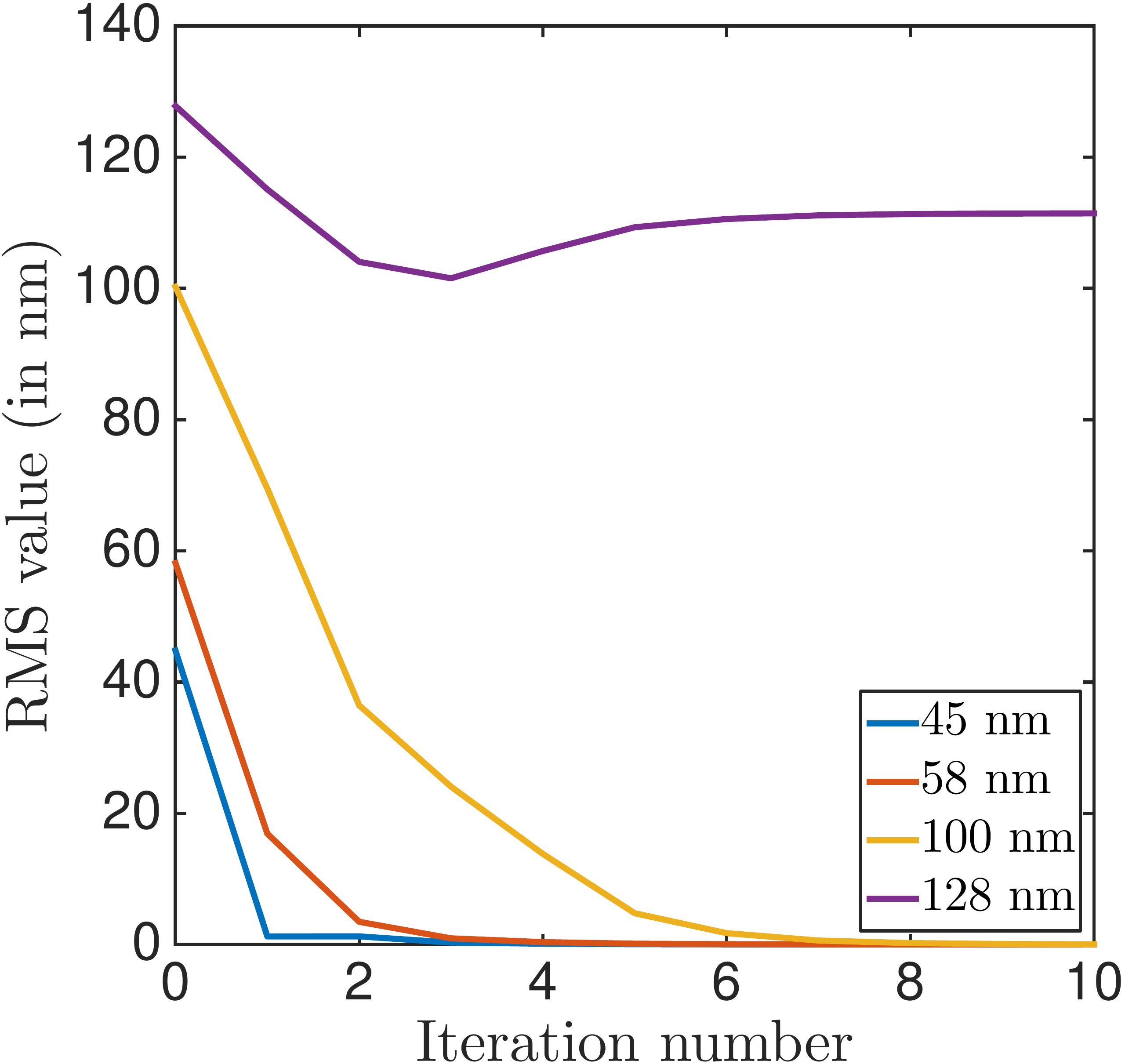}
\hspace{0.08\textwidth}
\includegraphics[height=0.38\textwidth]{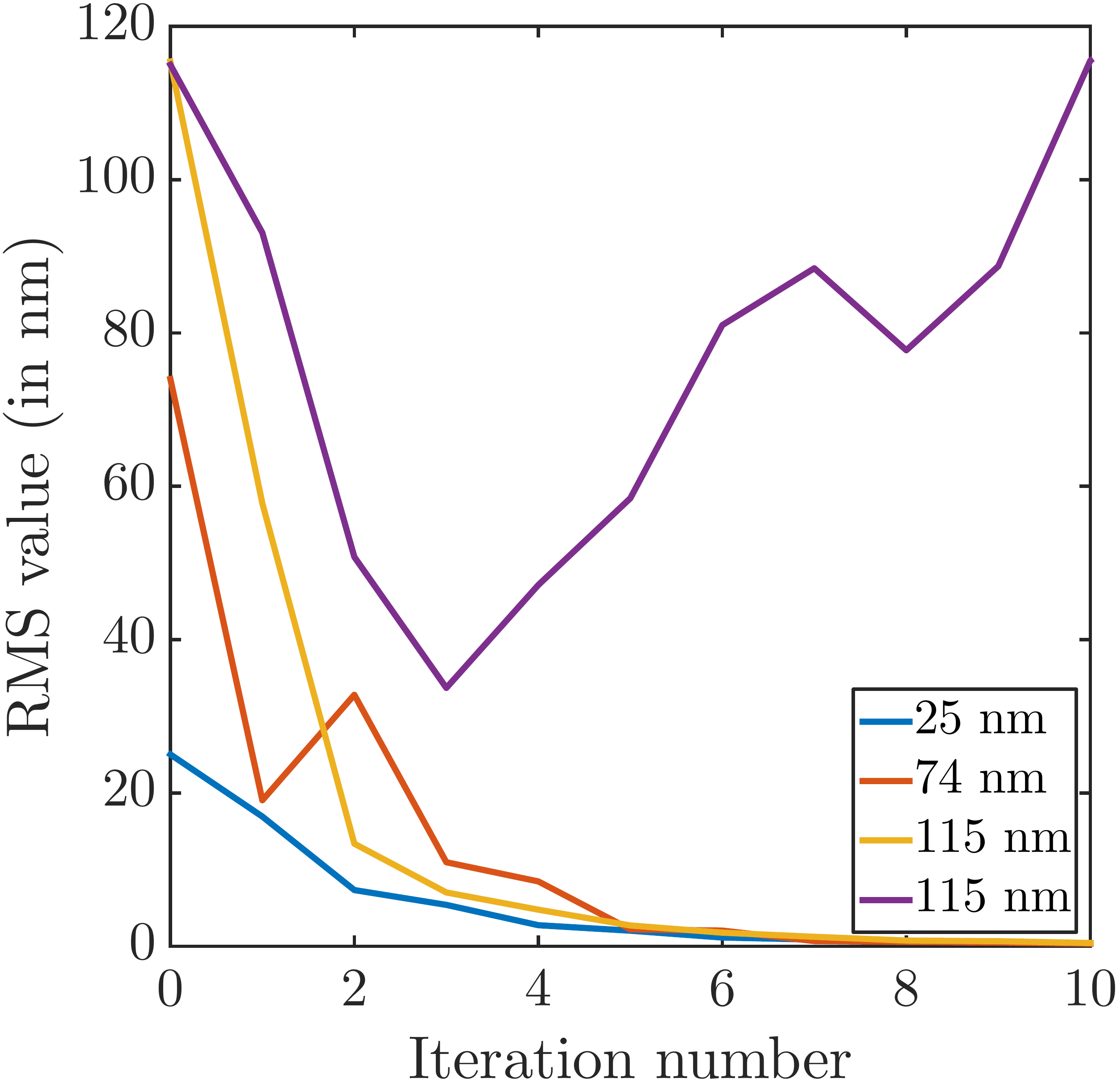}
\caption{RMS value evolution of the aberrations over the pupil as a function of the iteration number in the case of piston aberrations only (left) and for piston and tip-tilt (right). The starting RMS values are given in the captions.}
\label{ConvergencePiston}
\end{figure*}
By using a simple rule of thumb assuming the Nyquist-Shannon sampling theorem, a condition on $\gamma$ can be expressed as
\begin{equation}
\gamma \ge 4N+2,
\label{gamma}
\end{equation}
where $N$ represents the number of segment rings over the telescope pupil.
As we observe in Eq. \ref{gamma}, the higher the number of segments, or rings ($N$), the higher the value of $\gamma$. 
Conversely, the higher the value of $\gamma$, the lower the fringes contrast in the pupil plane. 
Assuming 18 segments over 2 rings as for the JWST, or 36 segments over 3 rings as for the Keck telescope, leads to $\gamma \ge 10$ and $\gamma \ge 14,$ respectively. Whereas these conditions are reasonable to consider, the condition that must satisfy an implementation on an ELT (e.g., for the 798 segments over 16 rings of the E-ELT, the condition is $\gamma \ge 66$) is unrealistic as it consequently leads to a very poor S/N. \\
At first glance, it precludes the operation of the SCC-PS on an ELT, but a solution exists to improve the S/N associated with a given $\gamma$, even for very high $\gamma$ values. The diffracted light distribution in the Lyot plane directly depends on the focal plane mask architecture installed in the SCC-PS system. 
Former and more recent studies have demonstrated that adequate focal plane mask designs allow a control of the energy spatial distribution outside the geometrical pupil \citep[e.g.,][]{SOUMMER03, NEWMAN15}, improving the level of energy, which falls into the reference channel and, consequently, its related S/N.

\section{Performance and discussions}
\label{sec:Results}

To assess the performance of the SCC-PS for retrieving and disentangling piston and tip-tilt information, we carried out the following series of tests:
(1) a closed-loop performance evaluation; (2) a sensitivity analysis to optical components misalignment; and (3) a sky coverage evaluation.

\begin{figure}[!ht]
\centering
\includegraphics[height=0.38\textwidth]{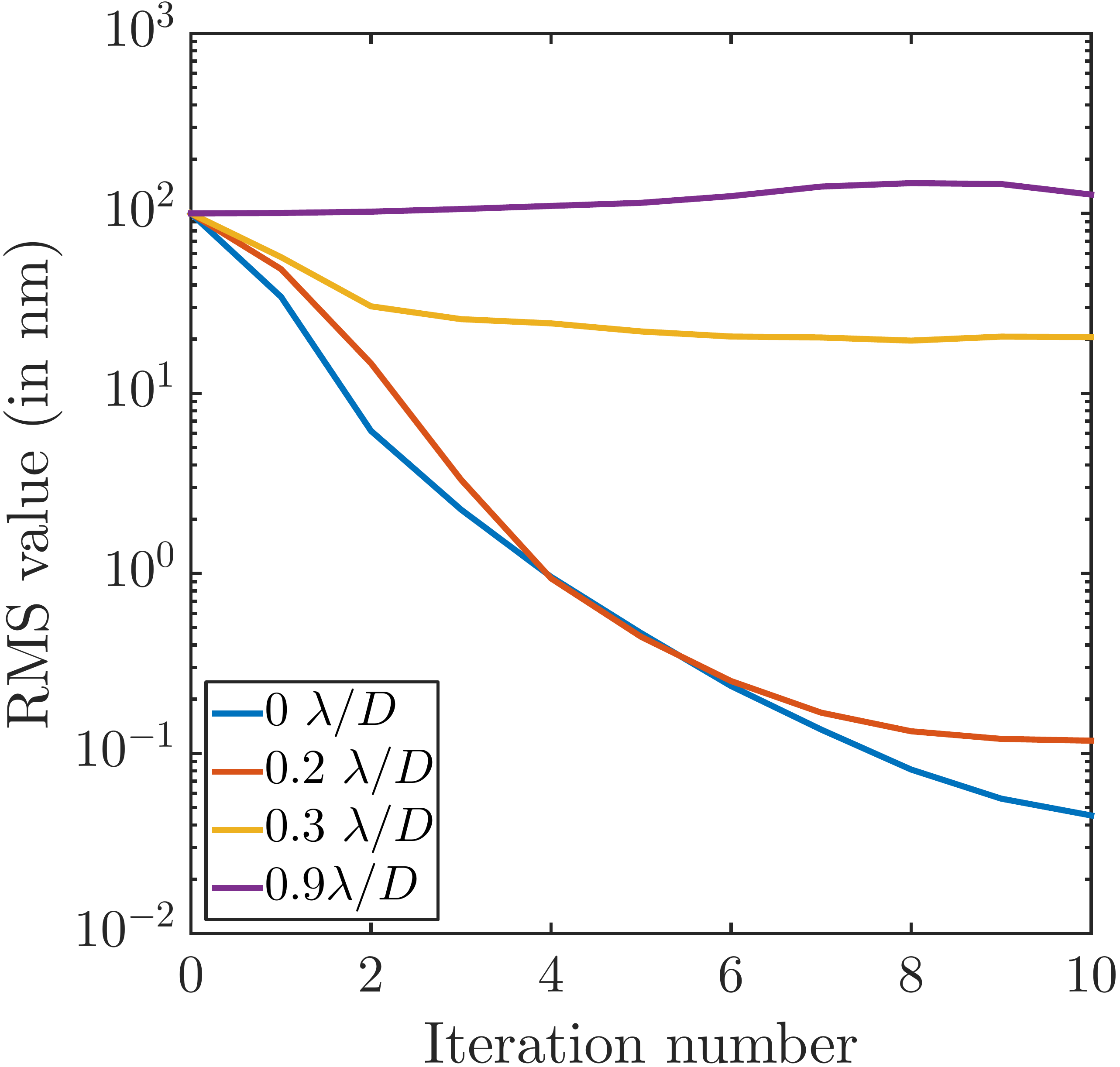}
\caption{Impact of the focal plane mask misalignment (expressed in $\lambda/D$) in the estimation of the residual RMS over the pupil as a function of the iteration number.}
\label{Align1}
\end{figure} 
\subsection{Closed-loop accuracy}
\label{subsec:closeloop}

We independently analyze the convergence and the residual RMS in two cases: first with piston only and then with both piston and tip-tilt. 
Figure \ref{ConvergencePiston} (left) presents the residual aberrations measured on the mirror as a function of the iteration number for various initial piston only configurations, whereas Fig. \ref{ConvergencePiston} (right) shows the residual RMS as a function of the iteration number for various initial combinations of piston and tip-tilt RMS. 
Dedicated calibration matrix are used for each case. For piston simulations, segments are poked only with a piston while $\varphi_0$  is only recorded. For piston and tip-tilt simulations, it is built following the method presented in Sec. \ref{subsec:calibration}.
In both cases and as expected the results of the phasing strictly depends on the initial configuration of the pupil.

\begin{figure}[!ht]
\centering
\includegraphics[height=0.38\textwidth]{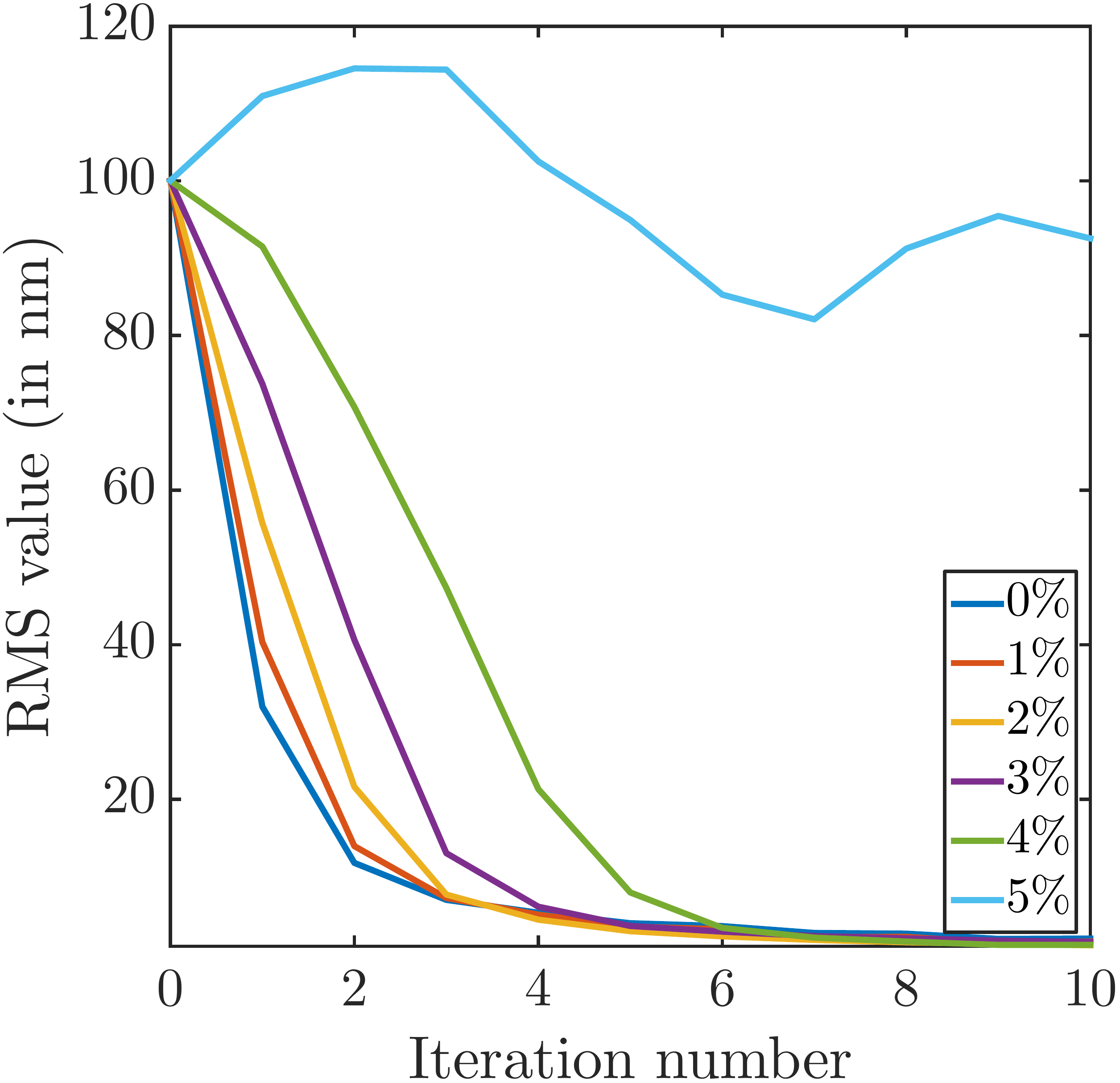}
\caption{RMS value as a function of the iteration number for various pupil shear amplitudes. The shifts are given in percentage of the pupil diameter for a JWST-like mirror ($N=2$).}
\label{DecalagePupille}
\end{figure} 
For piston only simulations (Fig. \ref{ConvergencePiston}, left), two outcomes are possible: (1) the correction converges to $0$ if the initial RMS is within the capture range of the SCC-PS; (2) the correction converges to a nonzero value (purple curve on Fig. \ref{ConvergencePiston}, left). \\
\noindent In case (1), the correction is ideal. The residual RMS is then nearly null ($\sim 7.10^{-2}$\,nm RMS) and the estimated Strehl ratio is roughly $100\%$. Quantitatively, it is remarkable that the convergence requires so few iterations. \\ 
\noindent Case (2) presents an effect known as the so-called $\lambda$-ambiguity, where part of the segments are positioned to a multiple of the wavelength away from the correct position. 
This effect is described well in the literature; \citet{VIGAN11} provides a pedagogic illustration and description of the phenomenon.
The $\lambda$-ambiguity effect is highly visible on the RMS values, but it leads to a Strehl ratio of roughly $100\%$ (in monochromatic light).

Simulations including both piston and tip-tilt (Fig. \ref{ConvergencePiston}, right) lead to similar ending scenarios. In case (1), the system converges to a perfectly phased pupil, with, as for piston only, nearly null residual RMS. 
In case (2), the system still converges to a nonzero RMS value (purple curve on Fig. \ref{ConvergencePiston}, right and for numbers of iterations $\sim 20$). However, the Strehl ratio is not close to $100\%$ anymore due to the presence of tip-tilt. Indeed, there are nonzero values of tip-tilt for which the SCC-PS estimates $\varphi_1$ or $\varphi_2$ as zero, thus leading the system to an inappropriately phased but steady state.
There is no clear limit in terms of starting RMS value between a converging and a nonconverging system. As shown in Fig. \ref{ConvergencePiston} (right), the yellow and purple curves start from the same RMS value but while the yellow curve is converging toward zero, the purple curve is not. The aberrations introduced over the initial pupil are randomly determined following a normal distribution and thus the convergence depends on whether or not piston and tip-tilt amplitudes of all the segments of the pupil lie in the capture range of the sensor. It means that for two identical values of initial RMS over the pupil, one pupil can have all its segments in the capture range of the SCC-PS while the second does not satisfy these requirements for a few segments.

For the sake of a comparison, we confronted our results to those obtained by \citet{YAITSKOVA04} using a Mach-Zehnder interferometer phasing sensor in a monochromatic regime at $500$\,nm. 
The SCC-PS demonstrates an improvement in accuracy and convergence speed. For instance, considering an initial RMS of $\sim \lambda/10$, the SCC-PS requires two iterations while the Mach-Zehnder interferometer sensor requires five iterations (Fig. 16 of the paper). 

Finally, these initial investigations we conducted (Sect. \ref{Numerical} and \ref{subsec:closeloop}) highlight the fact that the SCC-PS behave similarly as any cophasing sensor (e.g., capture range and closed-loop operation) emphasized by a subnanometric accuracy and rapid convergence.

\begin{figure}[!ht]
\centering
\includegraphics[height=0.38\textwidth]{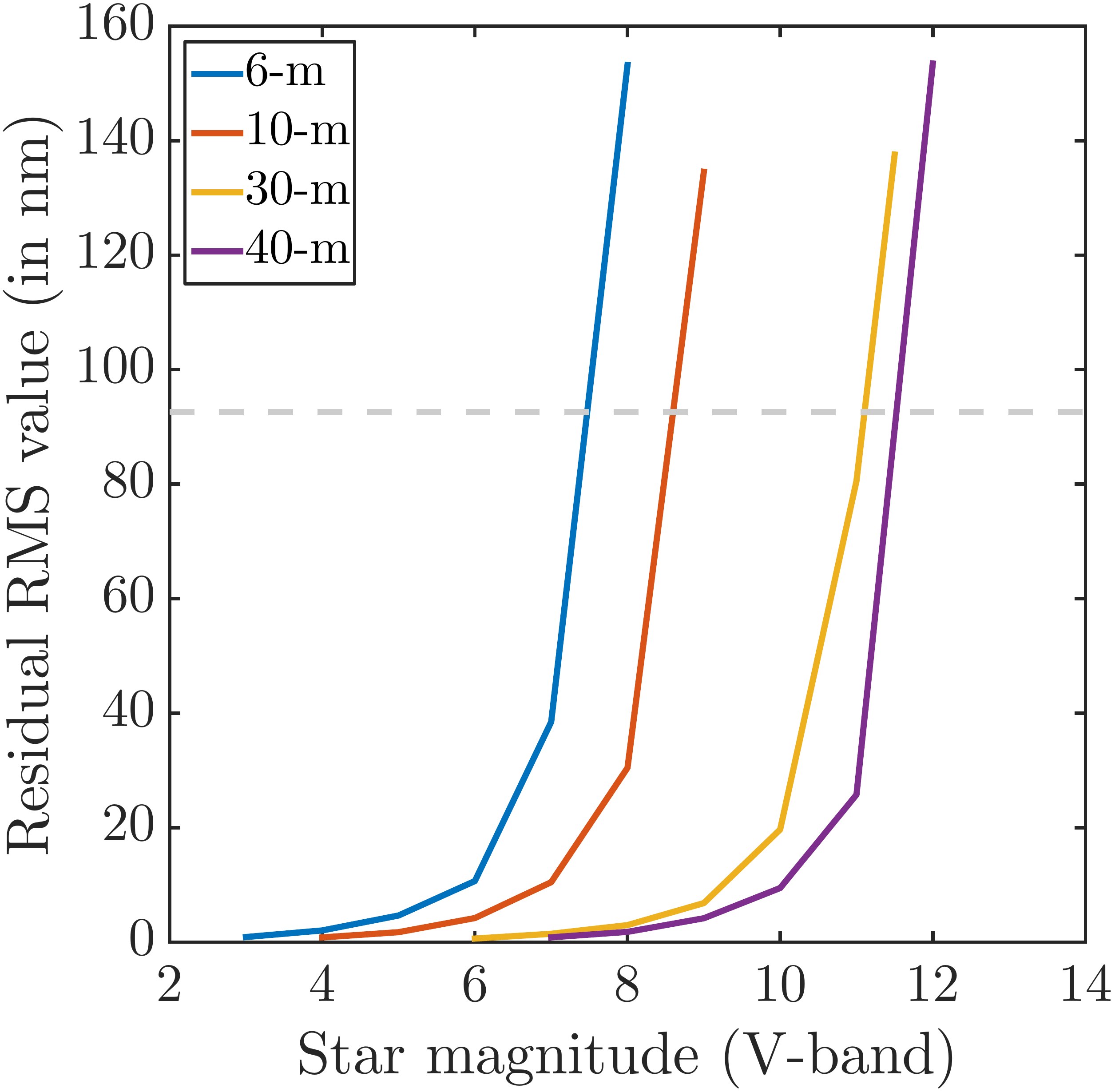}
\caption{ Residual nm rms as a function of the stellar magnitude ($V$ band) for various telescope diameters. The gray dashed line represents the initial wavefront error RMS introduced in the simulations. }
\label{Magnitude}
\end{figure} 
\subsection{Focal mask misalignment}

In this section, we arbitrarily assume 100 nm RMS initial wavefront error (piston and tip-tilt). This RMS value is high enough to be scientifically relevant and low enough to remain within the capture range of the sensor.
The objective of the following test is to appraise the sensitivity of the SCC-PS to the focal plane mask alignment. 
We use for this test a calibration matrix built from a system with a perfectly aligned FQPM.
Figure \ref{Align1} shows the impact of various amplitudes of the focal mask misalignments on the closed-loop convergence.
It is readily observable that departure from the ideal case (where the FQPM is perfectly aligned, blue curve) starts to be significant for misalignments higher than 0.2 $\lambda/D$.
For value of 0.3 $\lambda/D$ (yellow curve) the convergence reaches a threshold at basically 20 nm RMS, whereas for higher value of misalignment the system diverges (purple curve). 
 
The SCC-PS demonstrates a slight sensitivity to the focal plane mask alignment to levels that can be straightforwardly guaranteed. 
Indeed low-order wavefront sensor used in conjunction with a coronagraph have already demonstrated tiny levels of control in the centering of starlight onto a coronagraph at a $10^{-2}$ level at the Paris Observatory \citep{Mas2012} and $10^{-3}$ level at the Subaru telescope \citep{Singh2014}.

\subsection{Pupil shear}

During real observations the position of the pupil and more particularly the position of all the segments must be precisely known to ensure an optimal phasing of the pupil. 
The telescope pupil may undergo significant mismatch that could amount to more than a few $\%$ of its diameter. For example, the JWST is expected to deliver a pupil image for whose position is known at about 3-4$\%$ \citep{BOS2011}.
We study this effect in the simulations by shifting the segmented mirror in the pupil plane and running the SCC-PS correction using a calibration matrix obtained with a perfectly aligned system.
Figure \ref{DecalagePupille} presents the residual mirror RMS as a function of the iteration number for different pupil shear values. 
For the sake of a comparison, the mirror used for these simulations is composed of two rings ($N=2$) to reproduce the pupil of the JWST.
It is observable that the system converges for pupil shear as high as $4\%$ of the pupil diameter (corresponding to $20\%$ of the segment diameter). This is in line with the above-mentioned JWST requirement. 
Therefore the SCC-PS is able to withstand a large amount of pupil shear without any recalibration requirements. The convergence speed is marginally affected as it seems to require $\sim 3$ additional iterations to get to the same residual RMS (for $4\%$ pupil shear). 
In contrast to other techniques \citep[e.g.,][]{PINNA08, MAZZOLENI08, CHUECA08, SURDEJTHESIS}, the SCC-PS is a simple sensor to align and does not require the interaction matrix to be recalibrated on a regular basis.

\subsection{Final accuracy and star magnitude}

Virtually all existing segmented telescopes perform optical phasing from a weekly to monthly basis using relatively bright natural guide stars \citep[$3^{th}$ to $7^{th}$ $V$-band magnitude assuming 20 to 60\,s typical exposure time, e.g.,][]{CHANAN99, Chanan2000, PINNA08, VIGAN11}. In the mean time the control of the segment positions relies on the stability of the edge sensors. Delivering continuous closed-loop updates of the reference values of the edge sensors would require optical phasing measurements during operation that require the phasing sensor to cope with a fainter star. The objective of the following series of tests is to evaluate the final accuracy of the SCC-PS as a function of the flux and to establish the faintest star magnitude with which it can operate. 
Usually the limiting star magnitude is defined as the faintest stellar magnitude with which the sensor is capable of aligning the segmented mirror toward a given accuracy. However, since the expected final accuracy strictly depends on the telescope specifications or observing programs (the high-precision phasing requirements deemed appropriate for a high-contrast instrument is relatively uncertain), we define hereafter the limiting star magnitude as the faintest star magnitude with which the SCC-PS close-loop still converge.

\subsubsection{Limiting stellar magnitude}

In the following, stellar magnitudes are assumed in the $V$ band and the spectral bandwidth and exposure time are $\Delta \lambda$ = 90\,nm and 30\,s, respectively. The bandwidth is only specified to provide the number of photons in the pupil plane but simulations remain monochromatic. The SCC has already been successfully used with polychromatic light \citep{MAZOYER13,DELORME16} for bandwidth up to $\Delta\lambda \sim$ 80\,nm. System transmission is ideal (100\%). Photon noise, readout noise (1 $\mathrm{e^{-}}$ rms), and dark current (1 $\mathrm{e^{-}}$/s) are accounted for and the pupil is made of 91 segments ($\gamma$=24). The sampling in the focal plane is still $\sim 8$ pixels per $\lambda/D$.

The simulation compares the final precision obtained after convergence for various stellar magnitudes and various telescope diameters representative of existing (JWST, Keck, and Gran Telescopio Canarias) and future space- and ground-based observatories (LUVOIR, TMT, and E-ELT). The results are shown in Fig. \ref{Magnitude} where the achievable star magnitude is (as expected) correlated with the telescope diameter. 
The results indicate that if the illumination of the mirror is provided by bright stars ($V \le 6$) the SCC-PS converges regardless of the telescope diameter. The SCC-PS diverges for stars dimmer than magnitude 7, 8, 10, and 11, for telescopes of diameter 6\,m, 10\,m, 30\,m, and 40\,m, respectively. It is established that mirror phasing can be achieved with relatively bright stars with the SCC-PS as usually envisioned with any phasing sensor.

\subsubsection{Impact of the number of segments}

In most of the previous simulations we arbitrarily assume 91 segments over the pupil independent of the telescope diameter. We chose these segments for two practical reasons: (1) to have a high enough number of segments in the pupil to validate our algorithms for substantively complex configurations, and (2) to keep a reasonable computational time without neglecting the sampling frequency of our simulations.
However, the number of segments of the telescope has an impact on the sensor sky coverage. 
We found that the higher the number of segments the lower the S/N in each individual segment. Consequently the limiting magnitude of the SCC-PS depends on $N$ as analogous to that of the number of subapertures in a Shack-Hartmann sensor and its limiting sensitivity.\\
To investigate the impact of $N$ on the sensor limiting stellar magnitude, we estimate and compare the limiting star magnitude achieved for a 10-m class telescope as a function of the number of segment rings ($N$) present in the primary mirror. We study the evolution of $N$ from 2 to 5, where $N$=3 is representative of the Keck or GTC primary mirror and $N$=2 is similar to the JWST configuration. For each $N$ configuration considered, the optimal $\gamma$ value is set according to Eq. \ref{gamma}.
As expected, we found that the higher the $N$ value the fainter the limiting star magnitude. We found that the limiting star magnitude increases by $\sim 1$ magnitude when $N$ decreases by 1 ring. 
In particular, the limiting star magnitude increases from 8 to 11 when $N$ decreases from 5 to 2 ($\gamma$ decreases by a factor of 2, from 25 to 12). 
If the limiting star magnitude is defined as the faintest magnitude with which the SCC-PS is capable of aligning the mirror at the level of 10 nm RMS residual, the limiting star magnitude would evolve from 10 to 6 when $N$ increases from 2 to 5.

\subsubsection{Sky coverage improvement}

Operation with fainter stars can be achieved to some extent with: (1) wider broadband optical filter measurements and by increasing the exposure time, (2) a modification of the detector binning (binning 2$\times$2, while making sure that the fringes sampling is high enough, usually helps increase the limiting star magnitude by $\sim 1.5$ magnitude), and (3) a more suitable diffractive focal plane mask that maximizes the flux in the reference channel. The typical magnitude limits we see here (between 7 and 11, see the previous subsection) are significantly different from the $15^{th}$ magnitude we see in the state-of-the-art cophasing sensors.
The reason for this apparent discrepancy is that the coronagraphic effect of the FQPM drastically rejects star light. As a consequence, and because the reference channel has to be small enough (Sec. \ref{Sec:Reference}), only a small fraction of the star light is used to form fringes.
This explains why the SCC-PS is stacked to a magnitude that is basically below $V$=11, while most of cophasing sensor systems can accommodate magnitude up to $V$=15.
The optimization of a focal plane array to realize the best sensitivity by improving the energy spatial distribution in between the reference and pupil channels, and by emphasizing the equalization and maximization of fluxes, is a complex task of primary importance.

\section{Conclusion}
\label{conclusion}

Hitherto developed to discriminate a planet from a speckle \citep{BAUDOZ06} and later as a focal plane wavefront sensor correcting for phase and amplitude aberrations upstream of a coronagraph \citep{GALICHER08, MAZOYER13}, the adaptation of the SCC as a phasing sensor (SCC-PS) is demonstrated. The paper provides numerical evidence that mirror fine phasing is effective and possible with the SCC-PS. 

The SCC-PS offers various major advantages: direct measurement on the science image; limited amount of hardware (noninvasive method and no dedicated optical path); effective simultaneous estimates of piston and tip-tilt with subnanometric residuals ($\sim 7.10^{-2}$\,nm RMS); no a priori on the signal and in particular at the segment boundaries; insensitivity to gaps and segment edges; rapid convergence; and virtual insensitivity to focal plane mask or pupil misalignments. The latter advantage is of primary importance because pupil registration is critical for most of the actual phasing methods (e.g., modified Shack-Hartmann, Zernike sensor, etc.) as the subapertures or the signal analysis zone must be aligned accurately with respect to the intersegment edges. 

This new focal plane phasing method is applicable to several existing and near-future telescopes. 
In particular it can be implemented in two different manners upon science objectives and observing programs: (1) as part of the active optics system of a telescope for fine phasing, and (2) as part of the scientific instrument of a telescope for fine phasing and/or real-time stability control during observations. In both situations the SCC-PS could be implemented in a woofer/tweeter system architecture together with the conventional active optics phasing sensor (coarse phasing) of the telescope.
Furthermore, in situation (2) the SCC-PS can provide access to quasi-static aberrations (phase and amplitude) upstream of a coronagraph and segment misalignments (piston and tip-tilt) measurements.
The SCC-PS represents an opportunity of having two real-time sensors in one: a quasi-static aberration focal plane wavefront sensor (SCC) coupled with a fine cophasing focal plane sensor (SCC-PS).
This avoids multiple active and photon-hungry subsystems in an instrument as often seen as a significant shortcoming. 

Finally, optimal control of the energy spatial distribution outside the geometrical pupil using brand new diffractive mask designs \citep[e.g., ][]{NEWMAN15} to improve the level of energy that falls into the reference channel is a key aspect to tackle to improve the efficiency and operation domain, for instance for enhancing the sky coverage, of the sensor.

\begin{acknowledgements}
P. Janin-Potiron is grateful to Airbus Defense and Space (Toulouse, France) and the R\'egion PACA (Provence Alpes C\^ote d\'{}Azur, France) 2014 PhD program for supporting his PhD fellowship. Authors also want to thank F. Martinache for the discussions and helpful comments.
\end{acknowledgements}

\bibliography{biblio_SCC}

\begin{thebibliography}{35}
\expandafter\ifx\csname natexlab\endcsname\relax\def\natexlab#1{#1}\fi

\bibitem[{{Baudoz} {et~al.}(2006){Baudoz}, {Boccaletti}, {Baudrand}, \&
  {Rouan}}]{BAUDOZ06}
{Baudoz}, P., {Boccaletti}, A., {Baudrand}, J., \& {Rouan}, D. 2006, in IAU
  Colloq. 200: Direct Imaging of Exoplanets: Science and Techniques, ed.
  C.~{Aime} \& F.~{Vakili}, 553--558

\bibitem[{{Bos} {et~al.}(2011){Bos}, {Ohl}, \& {Kubalak}}]{BOS2011}
{Bos}, B.~J., {Ohl}, R.~G., \& {Kubalak}, D.~A. 2011, in Society of
  Photo-Optical Instrumentation Engineers (SPIE) Conference Series, Vol. 8131,
  Society of Photo-Optical Instrumentation Engineers (SPIE) Conference Series,
  0

\bibitem[{{Chanan} {et~al.}(1998){Chanan}, {Troy}, {Dekens}, {Michaels},
  {Nelson}, {Mast}, \& {Kirkman}}]{CHANAN98}
{Chanan}, G., {Troy}, M., {Dekens}, F., {et~al.} 1998, \ao, 37, 140

\bibitem[{{Chanan} {et~al.}(1999){Chanan}, {Troy}, \& {Sirko}}]{CHANAN99}
{Chanan}, G., {Troy}, M., \& {Sirko}, E. 1999, \ao, 38, 704

\bibitem[{{Chanan}(1989)}]{CHANAN89}
{Chanan}, G.~A. 1989, in Society of Photo-Optical Instrumentation Engineers
  (SPIE) Conference Series, Vol. 1036, Precision Instrument Design, ed. T.~C.
  {Bristow} \& A.~E. {Hatheway}, 59

\bibitem[{{Chanan} {et~al.}(2000{\natexlab{a}}){Chanan}, {Troy}, \&
  {Ohara}}]{CHANAN00}
{Chanan}, G.~A., {Troy}, M., \& {Ohara}, C.~M. 2000{\natexlab{a}}, in Society
  of Photo-Optical Instrumentation Engineers (SPIE) Conference Series, Vol.
  4003, Optical Design, Materials, Fabrication, and Maintenance, ed.
  P.~{Dierickx}, 188--202

\bibitem[{{Chanan} {et~al.}(2000{\natexlab{b}}){Chanan}, {Troy}, \&
  {Ohara}}]{Chanan2000}
{Chanan}, G.~A., {Troy}, M., \& {Ohara}, C.~M. 2000{\natexlab{b}}, in Society
  of Photo-Optical Instrumentation Engineers (SPIE) Conference Series, Vol.
  4003, Optical Design, Materials, Fabrication, and Maintenance, ed.
  P.~{Dierickx}, 188--202

\bibitem[{{Chueca} {et~al.}(2008){Chueca}, {Reyes}, {Schumacher}, \&
  {Montoya}}]{CHUECA08}
{Chueca}, S., {Reyes}, M., {Schumacher}, A., \& {Montoya}, L. 2008, in Society
  of Photo-Optical Instrumentation Engineers (SPIE) Conference Series, Vol.
  7012, Society of Photo-Optical Instrumentation Engineers (SPIE) Conference
  Series, 13

\bibitem[{{Codona} \& {Doble}(2015)}]{CODONA15}
{Codona}, J.~L. \& {Doble}, N. 2015, ArXiv e-prints

\bibitem[{{Cuevas} {et~al.}(2000){Cuevas}, {Orlov}, {Garfias}, {Voitsekhovich},
  \& {Sanchez}}]{CUEVAS00}
{Cuevas}, S., {Orlov}, V.~G., {Garfias}, F., {Voitsekhovich}, V.~V., \&
  {Sanchez}, L.~J. 2000, in Society of Photo-Optical Instrumentation Engineers
  (SPIE) Conference Series, Vol. 4003, Optical Design, Materials, Fabrication,
  and Maintenance, ed. P.~{Dierickx}, 291--302

\bibitem[{{Delavaquerie} {et~al.}(2010){Delavaquerie}, {Cassaing}, \&
  {Amans}}]{DELAV10}
{Delavaquerie}, E., {Cassaing}, F., \& {Amans}, J.-P. 2010, in Adaptative
  Optics for Extremely Large Telescopes, 5018

\bibitem[{{Delorme} {et~al.}(2016){Delorme}, {Galicher}, {Baudoz}, {Rousset},
  {Mazoyer}, \& {Dupuis}}]{DELORME16}
{Delorme}, J.~R., {Galicher}, R., {Baudoz}, P., {et~al.} 2016, \aap

\bibitem[{{Dohlen} {et~al.}(2006){Dohlen}, {Langlois}, {Lanzoni}, {Mazzanti},
  {Vigan}, {Montoya}, {Hernandez}, {Reyes}, {Surdej}, \&
  {Yaitskova}}]{DOHLEN06}
{Dohlen}, K., {Langlois}, M., {Lanzoni}, P., {et~al.} 2006, in Society of
  Photo-Optical Instrumentation Engineers (SPIE) Conference Series, Vol. 6267,
  Society of Photo-Optical Instrumentation Engineers (SPIE) Conference Series,
  34

\bibitem[{{Esposito} {et~al.}(2005){Esposito}, {Pinna}, {Puglisi}, {Tozzi}, \&
  {Stefanini}}]{ESPOSITO05}
{Esposito}, S., {Pinna}, E., {Puglisi}, A., {Tozzi}, A., \& {Stefanini}, P.
  2005, Optics Letters, 30, 2572

\bibitem[{{Galicher} {et~al.}(2008){Galicher}, {Baudoz}, \&
  {Rousset}}]{GALICHER08}
{Galicher}, R., {Baudoz}, P., \& {Rousset}, G. 2008, \aap, 488, L9

\bibitem[{{Galicher} {et~al.}(2010){Galicher}, {Baudoz}, {Rousset}, {Totems},
  \& {Mas}}]{GALICHER10}
{Galicher}, R., {Baudoz}, P., {Rousset}, G., {Totems}, J., \& {Mas}, M. 2010,
  \aap, 509, A31

\bibitem[{{Lofdahl} {et~al.}(1998){Lofdahl}, {Kendrick}, {Harwit}, {Mitchell},
  {Duncan}, {Seldin}, {Paxman}, \& {Acton}}]{LOFDAHL98}
{Lofdahl}, M.~G., {Kendrick}, R.~L., {Harwit}, A., {et~al.} 1998, in Society of
  Photo-Optical Instrumentation Engineers (SPIE) Conference Series, Vol. 3356,
  Space Telescopes and Instruments V, ed. P.~Y. {Bely} \& J.~B. {Breckinridge},
  1190--1201

\bibitem[{{Martinache}(2013)}]{MARTINACHE13}
{Martinache}, F. 2013, \pasp, 125, 422

\bibitem[{{Mas} {et~al.}(2012){Mas}, {Baudoz}, {Rousset}, \&
  {Galicher}}]{Mas2012}
{Mas}, M., {Baudoz}, P., {Rousset}, G., \& {Galicher}, R. 2012, \aap, 539, A126

\bibitem[{{Mazoyer} {et~al.}(2013){Mazoyer}, {Baudoz}, {Galicher}, {Mas}, \&
  {Rousset}}]{MAZOYER13}
{Mazoyer}, J., {Baudoz}, P., {Galicher}, R., {Mas}, M., \& {Rousset}, G. 2013,
  \aap, 557, A9

\bibitem[{{Mazzoleni} {et~al.}(2008){Mazzoleni}, {Gont{\'e}}, {Surdej},
  {Araujo}, {Brast}, {Derie}, {Duhoux}, {Dupuy}, {Frank}, {Karban}, {Noethe},
  \& {Yaitskova}}]{MAZZOLENI08}
{Mazzoleni}, R., {Gont{\'e}}, F., {Surdej}, I., {et~al.} 2008, in Society of
  Photo-Optical Instrumentation Engineers (SPIE) Conference Series, Vol. 7012,
  Society of Photo-Optical Instrumentation Engineers (SPIE) Conference Series,
  3

\bibitem[{Montoya-Martinez(2004)}]{MONTOYATHESIS}
Montoya-Martinez, L. 2004, Theses

\bibitem[{Newman {et~al.}(2015)Newman, Guyon, Balasubramanian, Belikov,
  Jovanovic, Martinache, \& Wilson}]{NEWMAN15}
Newman, K., Guyon, O., Balasubramanian, K., {et~al.} 2015, Publications of the
  Astronomical Society of the Pacific, 127, pp. 437

\bibitem[{{Pinna} {et~al.}(2008){Pinna}, {Quiros-Pacheco}, {Esposito},
  {Puglisi}, \& {Stefanini}}]{PINNA08}
{Pinna}, E., {Quiros-Pacheco}, F., {Esposito}, S., {Puglisi}, A., \&
  {Stefanini}, P. 2008, in Society of Photo-Optical Instrumentation Engineers
  (SPIE) Conference Series, Vol. 7012, Society of Photo-Optical Instrumentation
  Engineers (SPIE) Conference Series, 3

\bibitem[{{Pope} {et~al.}(2014){Pope}, {Cvetojevic}, {Cheetham}, {Martinache},
  {Norris}, \& {Tuthill}}]{POPE2014}
{Pope}, B., {Cvetojevic}, N., {Cheetham}, A., {et~al.} 2014, \mnras, 440, 125

\bibitem[{{Rouan} {et~al.}(2000){Rouan}, {Riaud}, {Boccaletti}, {Cl{\'e}net},
  \& {Labeyrie}}]{ROUAN00}
{Rouan}, D., {Riaud}, P., {Boccaletti}, A., {Cl{\'e}net}, Y., \& {Labeyrie}, A.
  2000, \pasp, 112, 1479

\bibitem[{{Singh} {et~al.}(2014){Singh}, {Martinache}, {Baudoz}, {Guyon},
  {Matsuo}, {Jovanovic}, \& {Clergeon}}]{Singh2014}
{Singh}, G., {Martinache}, F., {Baudoz}, P., {et~al.} 2014, \pasp, 126, 586

\bibitem[{{Soummer} {et~al.}(2003){Soummer}, {Dohlen}, \& {Aime}}]{SOUMMER03}
{Soummer}, R., {Dohlen}, K., \& {Aime}, C. 2003, \aap, 403, 369

\bibitem[{Surdej(2011)}]{SURDEJTHESIS}
Surdej, I. 2011, Theses

\bibitem[{{Surdej} {et~al.}(2010){Surdej}, {Yaitskova}, \& {Gonte}}]{SURDEJ10}
{Surdej}, I., {Yaitskova}, N., \& {Gonte}, F. 2010, \ao, 49, 4052

\bibitem[{{Troy} {et~al.}(2006){Troy}, {Crossfield}, {Chanan}, {Dumont},
  {Green}, \& {Macintosh}}]{TROY06}
{Troy}, M., {Crossfield}, I., {Chanan}, G., {et~al.} 2006, in Society of
  Photo-Optical Instrumentation Engineers (SPIE) Conference Series, Vol. 6272,
  Society of Photo-Optical Instrumentation Engineers (SPIE) Conference Series,
  2

\bibitem[{{Vigan} {et~al.}(2011){Vigan}, {Dohlen}, \& {Mazzanti}}]{VIGAN11}
{Vigan}, A., {Dohlen}, K., \& {Mazzanti}, S. 2011, \ao, 50, 2708

\bibitem[{{Yaitskova} {et~al.}(2003){Yaitskova}, {Dohlen}, \&
  {Dierickx}}]{YAITSKOVA03}
{Yaitskova}, N., {Dohlen}, K., \& {Dierickx}, P. 2003, Journal of the Optical
  Society of America A, 20, 1563

\bibitem[{{Yaitskova} {et~al.}(2005){Yaitskova}, {Dohlen}, {Dierickx}, \&
  {Montoya}}]{YAITSKOVA05}
{Yaitskova}, N., {Dohlen}, K., {Dierickx}, P., \& {Montoya}, L. 2005, Journal
  of the Optical Society of America A, 22, 1093

\bibitem[{{Yaitskova} {et~al.}(2004){Yaitskova}, {Montoya-Martinez}, {Dohlen},
  \& {Dierickx}}]{YAITSKOVA04}
{Yaitskova}, N., {Montoya-Martinez}, L., {Dohlen}, K., \& {Dierickx}, P. 2004,
  in Society of Photo-Optical Instrumentation Engineers (SPIE) Conference
  Series, Vol. 5489, Ground-based Telescopes, ed. J.~M. {Oschmann}, Jr.,
  1139--1151

\end{thebibliography}

\end{document}